# Irregularly Shaped γ´-Fe$_4$N Nanoparticles for Hyperthermia Treatment and T$_2$ Contrast-Enhanced Magnetic Resonance Imaging with Minimum Dose

Kai Wu†,⊥,*, Jinming Liu†,⊥, Renata Saha†,⊥, Bin Ma†, Diqing Su‡, Chaoyi Peng†, Jiajia Sun†,§, and Jian-Ping Wang†,*

†Department of Electrical and Computer Engineering, University of Minnesota, Minneapolis, Minnesota 55455, USA

‡Department of Chemical Engineering and Material Science, University of Minnesota, Minneapolis, Minnesota 55455, USA

§State Key Laboratory of Electrical Insulation and Power Equipment, Xi'an Jiaotong University, Xi'an, Shaanxi Province 710049, China

Abstract: Magnetic nanoparticles (MNPs) have been extensively used in drug/gene delivery, hyperthermia therapy, magnetic particle imaging (MPI), magnetic resonance imaging (MRI), magnetic bioassays, etc. With proper surface chemical modifications, physicochemically stable and non-toxic MNPs are emerging contrast agents and tracers for in vivo MRI and MPI applications. Herein, we report the high magnetic moment, irregularly shaped γ´-Fe$_4$N nanoparticles for enhanced hyperthermia therapy and T$_2$ contrast agent for MRI application. The static and dynamic magnetic properties of γ´-Fe$_4$N nanoparticles are characterized by vibrating sample magnetometer (VSM) and magnetic particle spectroscopy (MPS) systems, respectively. Compared to the γ-Fe$_2$O$_3$ nanoparticles, γ´-Fe$_4$N show at least 3 times higher saturation magnetization (in emu/g), which, as a result, gives rise to the stronger dynamic magnetic responses as proved in the MPS measurement results. In addition, γ´-Fe$_4$N nanoparticles are functionalized with oleic acid layer by a wet mechanical milling process, the morphologies of as-milled nanoparticles are characterized by transmission electron microscopy (TEM), dynamic light scattering (DLS) and nanoparticle tracking analyzer (NTA). We report that with proper surface chemical modification and tuning on morphologies, γ´-Fe$_4$N nanoparticles could be used as tiny heating sources for hyperthermia and contrast agents for MRI applications with minimum dose.

## 1. INTRODUCTION

Nowadays, magnetic nanoparticles (MNPs), with proper surface chemical modifications, are emerging nanomaterials that have been exploited in the areas of magnetic resonance imaging (MRI) and magnetic particle imaging (MPI),[1–10] drug/gene delivery,[11–16] hyperthermia,[17–25] bioassays,[10,26,27,27–30] cell sorting and separation,[31–35] etc. For different applications, high magnetic moment MNPs are demanded for larger magnetic torques in drug/gene delivery and cell sorting and separation applications, for high sensitivity magnetic bioassays, for efficient and minimum dose usage in MRI, MPI and hyperthermia applications. In view of this demand, $\gamma'$-$Fe_4N$ nanoparticles are reported as magnetically soft,[36] chemically stable,[37] cheap with high saturation magnetization (182 emu/g).[37,38] Since 2000, many groups have reported the facile fabrication of high purity $\gamma'$-$Fe_4N$ nanoparticles.[37–39] Based on the Fe-N phase diagram, $\gamma'$-$Fe_4N$ phase forms at the temperature range from 200 to 680 °C.[40] As proposed by the Lehrer diagram, the most stable iron nitride phase could be modified by tuning the nitriding potential that is controlled by the partial pressure of hydrogen and ammonia gas and the nitridation temperature.[41] Thus, different iron nitride phases are obtained using the gas nitridation process, such as $\alpha''$-$Fe_{16}N_2$, $\gamma'$-$Fe_4N$, $\gamma$-FeN, $\varepsilon$-$Fe_3N$, etc.[42–45] In this paper, we report the gas nitridation method to synthesize $\gamma'$-$Fe_4N$ nanoparticles. During the nitridation process, ammonia gas provides nitrogen atoms and hydrogen gas is applied to tune the nitriding potential to obtain $\gamma'$-$Fe_4N$ nanoparticles. Figure 1 shows the experimental setup for preparing $\gamma'$-$Fe_4N$ nanoparticles. Hydrogen gas is for reducing the starting materials, $\gamma$-$Fe_2O_3$ nanoparticles, in a tube furnace. Then the nitridation is proceeded in the same furnace under a mixture of ammonia and hydrogen gas to obtain $\gamma'$-$Fe_4N$ nanoparticles, as shown in Figure 1(a). The reduction and nitridation reactions are illustrated in Figure 1(b).

Herein, we report the fabrication of $\gamma'$-$Fe_4N$ nanoparticles from $\gamma$-$Fe_2O_3$ nanoparticles. The crystalline structures of both nanoparticles are characterized by X-ray diffraction (XRD) and high purity $\gamma'$-$Fe_4N$ phase is confirmed from our fabricated $\gamma'$-$Fe_4N$ nanoparticles. The $\gamma$-$Fe_2O_3$ and fabricated $\gamma'$-$Fe_4N$ powders are wet mechanically milled with oleic acid (OA) to functionalize nanoparticle surface with OA chemical groups and to effectively separate nanoparticles. The static and dynamic magnetic responses of $\gamma$-$Fe_2O_3$ and $\gamma'$-$Fe_4N$ nanoparticles in OA solution are measured and compared using the vibrating sample magnetometer (VSM) and magnetic particle spectroscopy (MPS) systems, respectively. In addition, the morphologies of $\gamma$-$Fe_2O_3$ and $\gamma'$-$Fe_4N$ nanoparticles are characterized by transmission electron microscopy (TEM). It is confirmed that $\gamma$-$Fe_2O_3$ nanoparticles, with average magnetic core size of 20 nm, are sintered into larger, irregularly shaped $\gamma'$-$Fe_4N$ nanoparticles of around 100 nm. The dynamic light scattering (DLS) and nanoparticle tracking analyzer (NTA) are used to measure the hydrodynamic size of both nanoparticles. The irregularly shaped $\gamma'$-$Fe_4N$ nanoparticles, with high saturation magnetizations, provide a new way for enhanced $T_2$ relaxivity in MRI and efficient hyperthermia treatment with minimum dose requirements.

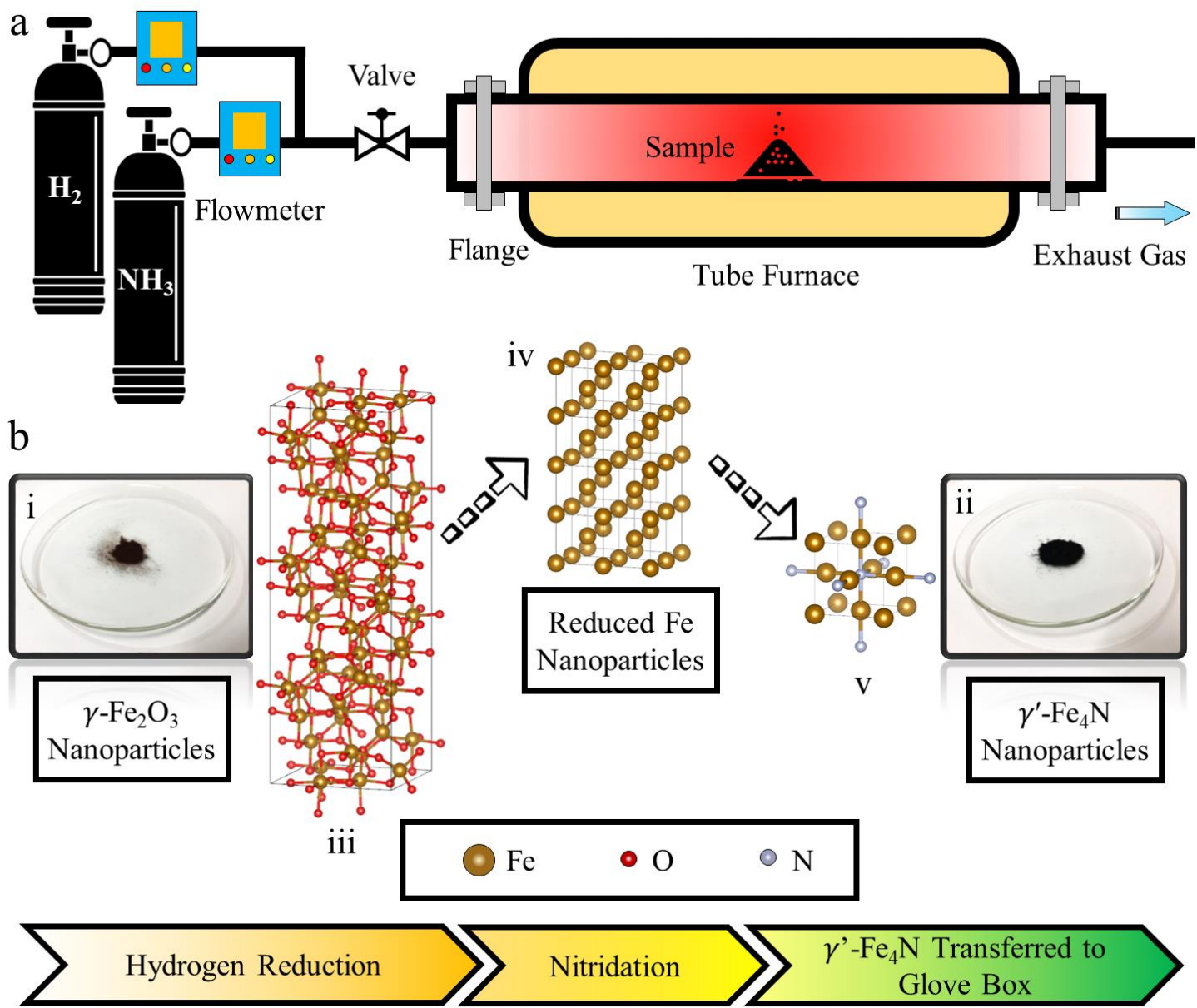

Figure 1. γ´-Fe$_4$N nanoparticles prepared by a gas nitridation approach. (a) The schematic drawing of the gas nitridation set up. The starting material γ-Fe$_2$O$_3$ nanoparticles are placed in a tube furnace. Hydrogen and ammonia gas cylinders provide high purity gas for the reduction and nitridation processes. (b) Summary on the working principle of gas nitridation. The γ-Fe$_2$O$_3$ nanoparticles, the starting materials, are reduced by hydrogen gas in a tube furnace. Then a mixture of hydrogen and ammonia are applied to synthesize γ´-Fe$_4$N nanoparticles. These synthesized γ´-Fe$_4$N nanoparticles are transferred into a glove box to avoid oxidation. (i) Photograph of γ-Fe$_2$O$_3$ powder, (ii) photograph of γ´-Fe$_4$N powder, (iii) crystal structure of γ-Fe$_2$O$_3$, (iv) crystal structure of α-Fe, (v) crystal structure of γ´-Fe$_4$N.

## 2. RESULTS AND DISCUSSION

**2.1. X-ray Diffraction (XRD) of γ-Fe$_2$O$_3$ and synthesized γ´-Fe$_4$N MNPs.** The structure of γ-Fe$_2$O$_3$ and fabricated γ´-Fe$_4$N are investigated by X-ray Diffraction (XRD). As shown in Figure 2(a), the XRD pattern of the starting material matches the γ-Fe$_2$O$_3$ phase. After the hydrogen reduction and gas nitridation, nanoparticles show that γ´-Fe$_4$N is the main phase. The γ´-Fe$_4$N nanoparticles are successfully synthesized by the gas nitridation method. There is also a diffraction peak at around 2 theta 36 degrees that is from iron oxide, which might be due to the oxidation when transferring powder sample from the tube furnace to a glove box. The crystal structure of γ´-Fe$_4$N and γ-Fe$_2$O$_3$ are also plotted in Figure 2(b) & (c), respectively.

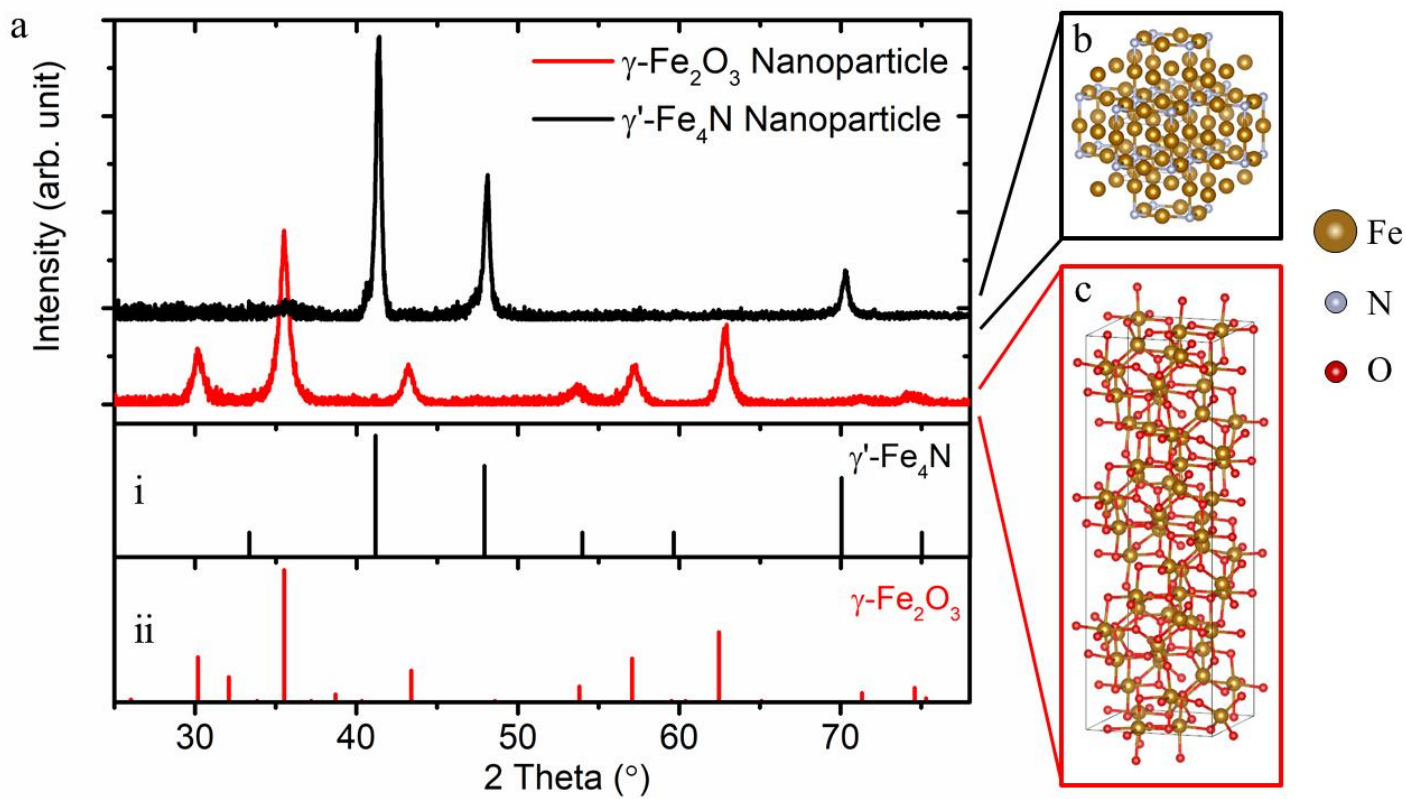

Figure 2. XRD patterns and crystal structures of γ-Fe$_2$O$_3$ and synthesized γ´-Fe$_4$N nanoparticles. (a) XRD patterns of γ-Fe$_2$O$_3$ nanoparticles (red solid line) and nitride nanoparticles (black solid line). The powder diffraction files (PDFs) are also plotted, as shown in the panels (i) γ´-Fe$_4$N and (ii) γ-Fe$_2$O$_3$. The XRD pattern of the starting material, iron oxide nanoparticles, matches well with the γ-Fe$_2$O$_3$ (PDF card No. 00-004-0755). Based on the PDF of γ´-Fe$_4$N (PDF card No. 00-006-0627), the main phase of the synthesized nitride nanoparticles is γ´-Fe$_4$N. A tiny iron oxide peak around 36 degree is observed from the synthesized nitride nanoparticle sample. The oxidation might happen during the sample transfer process such as from the tube to the glove box. (b) and (c) reveal the crystal structures of γ´-Fe$_4$N and γ-Fe$_2$O$_3$, respectively.

**2.2. Morphology characterization on γ-Fe2O3 and γ´-Fe4N nanoparticles.** The morphologies of γ-Fe$_2$O$_3$ and γ´-Fe$_4$N nanoparticles are obtained using a transmission electron microscopy (TEM). From each nanoparticle sample, three different samples are prepared for the TEM characterization: the wet mechanical milled γ-Fe$_2$O$_3$ and γ´-Fe$_4$N nanoparticles in oleic acid (OA), named as γ-Fe$_2$O$_3$@BM and γ´-Fe$_4$N@BM; the supernatant of these nanoparticle suspensions after a ultra-centrifugation step (10,000 rpm for 20 min), named as γ-Fe$_2$O$_3$@Ultra and γ´-Fe$_4$N@Ultra; the supernatant from these nanoparticle suspension after keeping at room temperature for 24 h, named as γ-Fe$_2$O$_3$@Sup and γ´-Fe$_4$N@Sup. The TEM images, illustration of each TEM sample preparation process, and schematic drawings of different shape nanoparticles are summarized in Figure 3(a) & (b). Both γ-Fe$_2$O$_3$@Ultra and γ´-Fe$_4$N@Ultra show well dispersed nanoparticles with diameters below 20 nm. Most of these nanoparticles have irregular shapes. TEM images of the as-milled nanoparticles, γ-Fe$_2$O$_3$@BM and γ´-Fe$_4$N@BM show that nanoparticles tend to aggregate together to minimize their surface energy. Around 60% of the nanoparticles from the γ´-Fe$_4$N@BM sample are sintered bodies and aggregate together, which might happen during the reduction and nitridation which are handled at a relatively high temperature (400 $_o$C). These sintered bodies are also observed from the γ´-Fe$_4$N@Sup sample. The γ-Fe$_2$O$_3$@Sup sample does not have any sintered bodies but aggregations are still observed. Various shapes of nanoparticles are highlighted in the TEM images shown in Figure 3 by dashed lines. The corresponding schematic shapes of single nanoparticles and sintered bodies are drew in Figure 3(i) - (viii). Micromagnetic simulations on the static magnetic responses of these γ-Fe$_2$O$_3$ and γ´-Fe$_4$N nanoparticles (sintered bodies) are given in Figure 6 and Supplementary Materials S5.

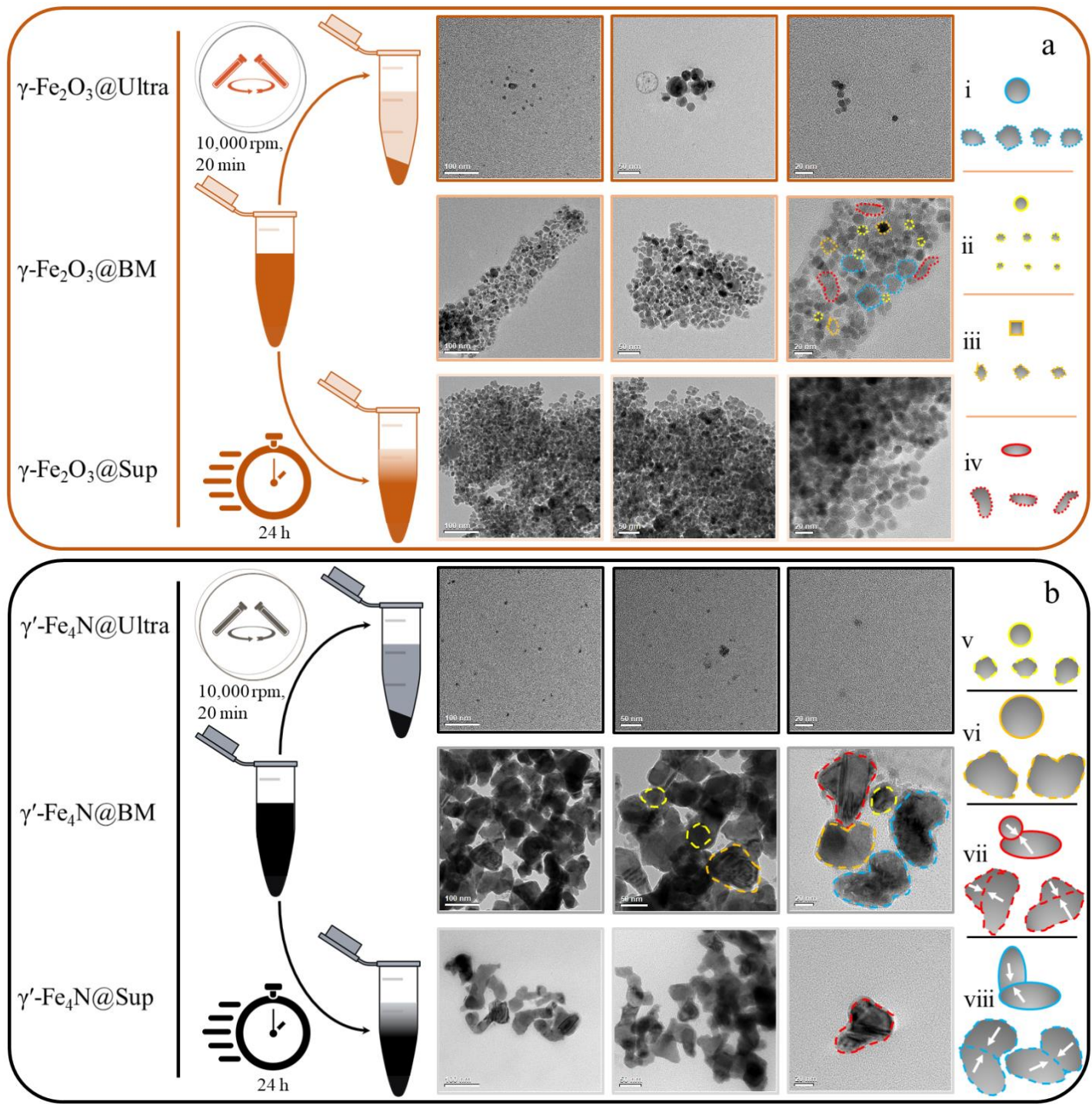

Figure 3. TEM sample preparation and bright field images of γ-Fe$_2$O$_3$ and γ´-Fe$_4$N nanoparticles. Three different samples are prepared for the TEM measurements: original nanoparticle suspensions of as-milled nanoparticles in OA are named as BM; original nanoparticle suspensions ultra-centrifuged at 10,000 rpm for 20 min are named as Ultra; original nanoparticle suspensions suspended for 24 h are named as Sup. A drop of each suspension is obtained from supernatant and dropped onto TEM grid (copper meshes with amorphous carbon coating) respectively. The TEM grids are air-dried at room temperature, which will be used for TEM measurements. (a) TEM characterization on γ-Fe$_2$O$_3$ nanoparticle samples. Three different samples are characterized: γ-

Fe$_2$O$_3$@Ultra, γ-Fe$_2$O$_3$@BM, and γ-Fe$_2$O$_3$@Sup. The corresponding TEM images are shown on the right panel. Most of the nanoparticles show irregular shapes with some spherical nanoparticles, as shown in (i), (ii), (iii), and (iv). Less aggregations of nanoparticles are observed from the supernatant of the sample γ-Fe$_2$O$_3$@Ulta. (b) TEM characterization on γ´-Fe$_4$N nanoparticle samples. Three different samples are characterized: γ´-Fe$_4$N@Ultra, γ´-Fe$_4$N@BM, and γ´-Fe$_4$N@Sup. Well-dispersed nanoparticles are observed from the γ´-Fe$_4$N@Ultra sample. For the samples γ´-Fe$_4$N@BM and γ´-Fe$_4$N@Sup, most of the nanoparticles show irregular shapes and the size of the nanoparticles are larger than that of the starting material, γ-Fe$_2$O$_3$ nanoparticles, as shown in (v), (vi), (vii), and (viii). More aggregations are observed which is due to the sintering of nanoparticles during the reduction and nitridation processes as well as the non-superparamagnetic properties of these sintered bodies.

**2.3. Hydrodynamic size of γ-Fe$_2$O$_3$ and γ´-Fe$_4$N nanoparticles.** Figure 4(a) shows the hydrodynamic size distribution of γ´-Fe$_2$O$_3$@BM sample measured by the dynamic light scattering (DLS), where the size distribution peaks around 25 nm and this too correlates with the TEM image Figure 4(i) added in the subset. In comparison to γ´-Fe$_4$N@BM, γ´-Fe$_2$O$_3$@BM sample shows far less aggregations and this suggests that they have formed well-dispersed nanoparticle sample solution in Isopar G fluid. Figure 4(b) shows the hydrodynamic size distribution of γ´-Fe$_4$N@BM sample where the peak is around 100 nm, agreeing with the TEM image Figure 4(ii) added in the subset. The DLS result of the γ´-Fe$_4$N@BM show another peak at around 1000 nm, this finding suggests that the γ´-Fe$_4$N MNPs are aggregated into clusters. A photograph of γ´-Fe$_4$N@BM sample is given in the glass bottle where the uppermost part is the supernatant while the sediments are nanoparticle clusters. Another interesting point to be noted is that the hydrodynamic size distribution in the DLS results appear slightly larger than the observed TEM images. The reason behind this being that the γ-Fe$_2$O$_3$@BM and γ´-Fe$_4$N@BM nanoparticles are formed as a result of wet ball milling in oleic acid. Thus, these nanoparticles are coated with a thin layer of oleic acid (around 2 nm thick) and this makes the hydrodynamic size distribution from DLS appears to be slightly larger than that observed from the TEM images. The schematic drawing of OA coated MNPs is given in Figure 4(c).

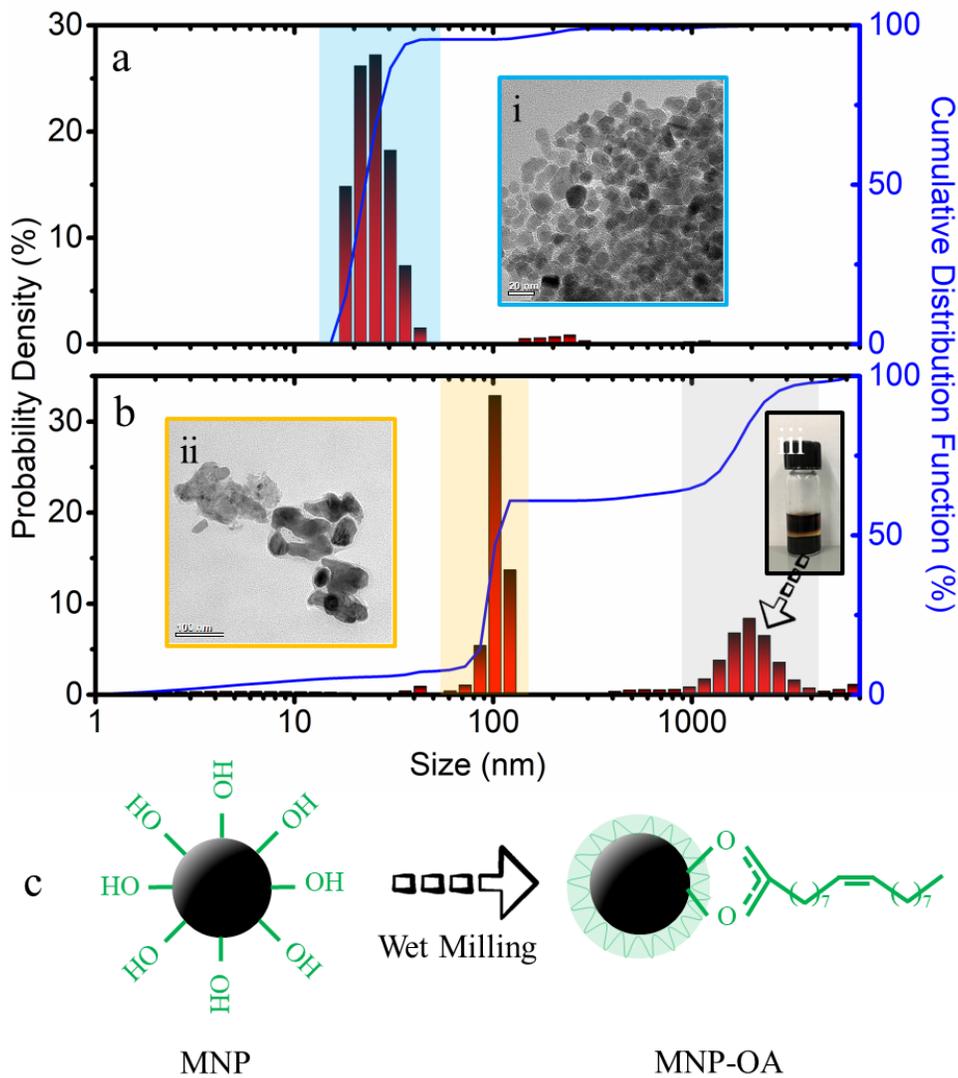

Figure 4. Hydrodynamic size distribution of (a) γ-Fe$_2$O$_3$@BM and (b) γ´-Fe$_4$N@BM samples measured by DLS. γ-Fe$_2$O$_3$@BM shows a peak of 25 nm which corroborates with the TEM image in the subset. In addition, it also shows very small peaks at > 100 nm suggesting very small amount of clusterings of the nanoparticles in the sample. γ´-Fe$_4$N@BM peaks around 100 nm implying significant sintering of the nanoparticles and the peaks near 1000 nm signifies the clustering of these sintered bodies. (c) A schematic drawing of OA surfactant conjugation on the nanoparticles. The thickness of oleic acid layer is around 2 nm, which increased the hydrodynamic size of nanoparticles by 4 nm.

**2.4. Hydrodynamic size and concentration of the γ´-Fe$_4$N@BM nanoparticles in Isopar G fluid.** The DLS results give a slight discrepancy with that of the TEM images. Although we justified the discrepancy by two explanations in the last section: first, as the samples were ball milled, they added an extra 2 nm oleic acid coating around the nanoparticles and made the hydrodynamic size distribution slightly larger than observed from the TEM images; second, the hydrodynamic peaks around 100 nm and 1000 nm from the γ´-Fe$_4$N@BM sample helped us infer that the particles might have sintered and aggregated. Thus, in order to further characterize the size of the fabricated γ´-Fe$_4$N nanoparticles, another competent optical characterization method, the Nanoparticle Tracking Analyzer (NTA) (details regarding the techniques and models have been mentioned in the Methods & Materials section), is used. An added advantage of characterizing the nanoparticles by the NTA over the DLS is that they also give an information about the concentration of the particles from the solution. Figure 5 shows that the γ´-Fe$_4$N@BM sample has a concentration of the order of 10$^7$ particles/mL. Five independent NTA measurements, each having a time span of 60 seconds, are carried out and labeled as curves I – V in Figure 5. The nanoparticle concentration is averaged over five measurements and represented by the shadowed curve in Figure 5. For each measurement, a 1-minute video is recorded by the camera of the NTA (videos are provided in the Supplementary Videos). Snapshots at the 15$_{th}$, 30$_{th}$, 45$_{th}$ and 60$_{th}$ seconds are summarized on the right panel of Figure 5. The NTA results also confirm that the fabricated γ´-Fe$_4$N nanoparticles show both sintering (peaks at 46 nm, 136 nm and 187 nm) as well as clustering (peaks at 609 nm and over 1000 nm) as was concluded from the DLS results. It is to be noted here that, the characterization of γ´-Fe$_4$N@BM samples by NTA have been reported and the same by γ-Fe$_2$O$_3$@BM have not been made. This argument has been addressed in the Materials & Methods Section of this paper.

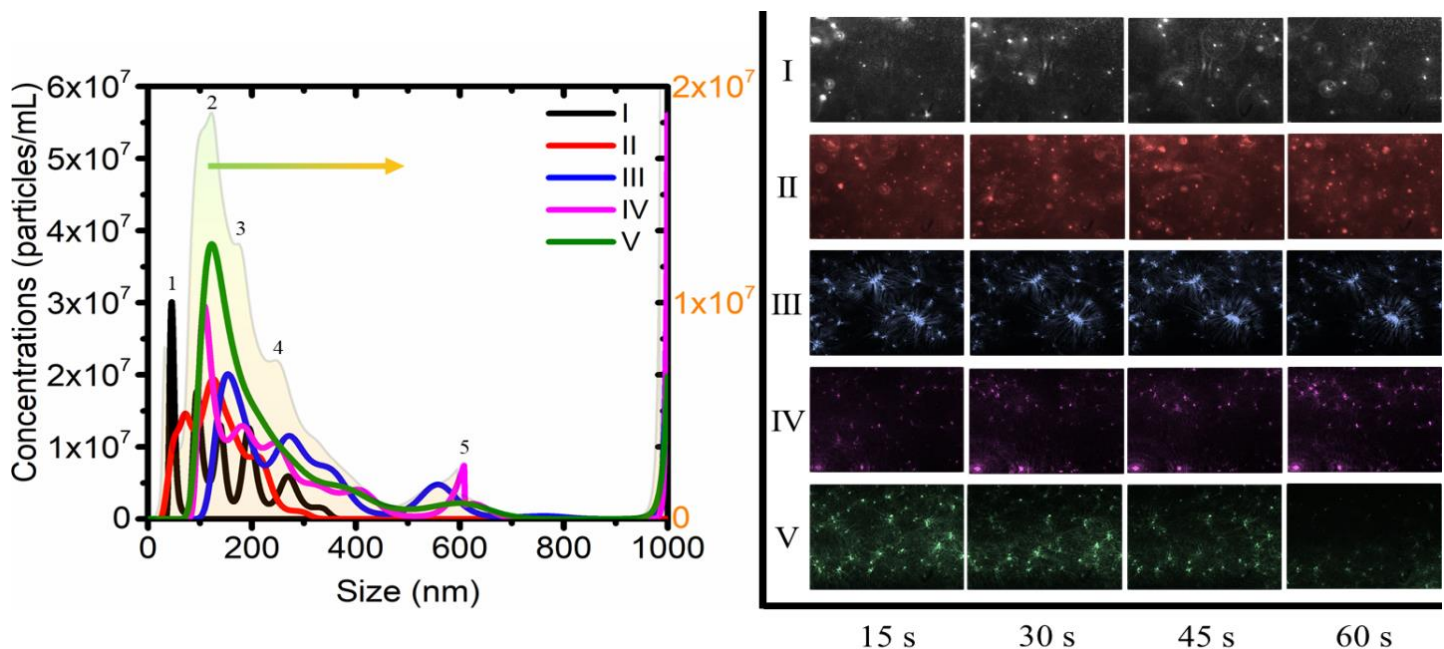

Figure 5. The hydrodynamic size distribution of γ´-Fe$_4$N@BM nanoparticles measured by the Nanoparticle Tracking Analyzer (NTA). Five independent runs are carried out on the sample. The snapshots of the γ´-Fe$_4$N@BM sample collected under the 403 nm wavelength light projected under the microscope for the NTA. The background color-codes of the snapshots corresponds to the color codes of the hydrodynamic size distribution curves from the 5 consecutive runs, each of 60 seconds at the 15$_{th}$, 30$_{th}$, 45$_{th}$ and 60$_{th}$ second. Bright spots from the snapshots are nanoparticles with larger spots representing clustered nanoparticles. The NTA size distribution corroborates well with DLS results of the γ´-Fe$_4$N@BM as discussed in Figure 4(b).

**2.5. Static magnetization curves of γ-Fe2O3 and γ´-Fe4N nanoparticles measured by vibrating sample magnetometer (VSM).** The static (DC) hysteresis loops of γ-Fe2O3@BM and γ´-Fe4N@BM samples are measured at room temperature by a VSM system, as plotted in Figure 6(a) & (b). The magnetic field is swept from -5000 to +5000 Oe with a step width of 5 Oe (or -2000 to +2000 Oe with a step width of 2 Oe), the averaging time for each data point is 100 ms so that the magnetizations of nanoparticles are able to relax and align to the field direction. The γ-Fe2O3@BM nanoparticles show a saturation magnetization $M_s$ of 51 emu/g and superparamagnetic property with negligible coercivity of 22 Oe. On the other hand, the γ´-Fe4N@BM nanoparticles show a hysteresis loop with a coercivity of 166 Oe and saturation magnetization $M_s$ of 164 emu/g. This large coercivity explains the sintered bodies (sizes around 100 nm) from the TEM, DLS, and NTA results in Figures 3(b), 4(b), and 5, respectively. At zero field, the remanent magnetization is around $26\% M_s$, which explains the severe aggregations from γ´-Fe4N@BM sample in Figures 4(b) and 5. Due to this non-superparamagnetic property and sintered bodies of fabricated γ´-Fe4N nanoparticles, the stability of γ´-Fe4N nanoparticles in OA is examined and given in the Supplementary Materials S7. Where the γ´-Fe4N@BM, γ-Fe2O3@BM, γ´-Fe4N@Ultra, γ-Fe2O3@Ultra samples are placed at room temperature for 7 days' of continuous observations.

**2.6. Mumax3 simulation and magnetic properties analysis.** Micromagnetic simulations conforming the different nanoparticle shapes observed in Figure 3(i) – (viii) are carried out on Mumax3. The simulation results show that all the γ-Fe2O3 nanoparticles modeled in this work are superparamagnetic and their magnetic moments align to the external DC field as marco-spins. On the other hand, the 100 nm γ´-Fe4N nanoparticle show domain walls and its remanent magnetization is non-negligible (see the magnetization in Figure 6(c): S-e at 0 Oe). In addition, the nanoparticle clusters are also simulated in this work (γ´-Fe4N nanoparticles with average size of above 500 nm from Figure 4(b) and Figure 5) as shown in Figure 6(c): S-g. Where the remanent magnetization is very significant, which contributes to the hysteresis loop in the VSM result.

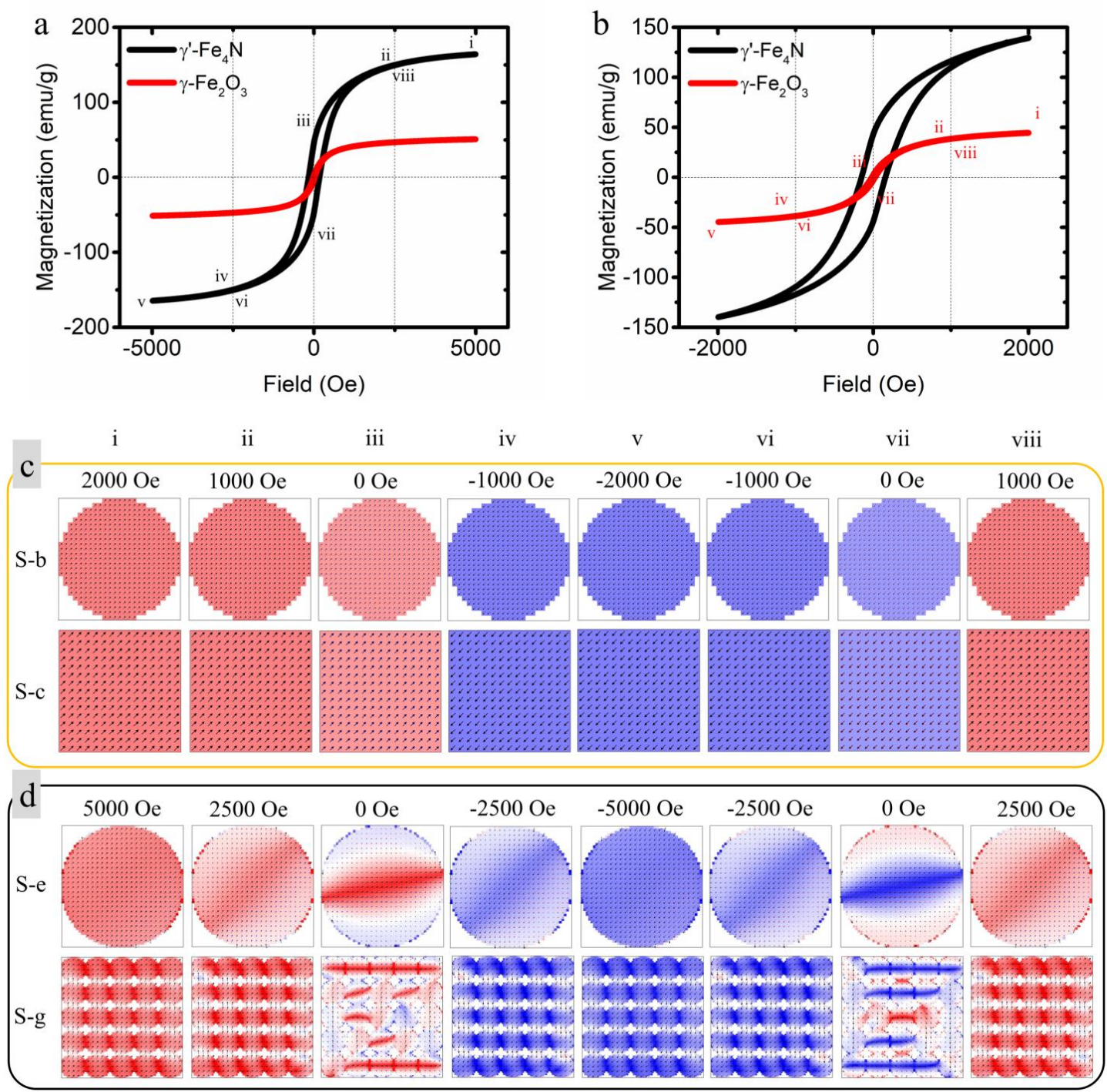

Figure 6. The DC (static) magnetization curves of dried γ-Fe$_2$O$_3$ and γ´-Fe$_4$N nanoparticles measured by VSM under external field ranging from (a) -5000 to 5000 Oe and (b) -2000 to 2000 Oe. The γ-Fe$_2$O$_3$ nanoparticles saturate at 2000 Oe field with a saturation magnetization of 51 emu/g and coercivity of 22 Oe. The γ´-Fe$_4$N nanoparticles saturate at 4000 Oe field with a saturation magnetization of 164 emu/g and coercivity of 166 Oe. Evolution of magnetizations in different (c) γ-Fe$_2$O$_3$ and (d) γ´-Fe$_4$N nanoparticles under different DC magnetic fields. S-b, S-c, S-e, and S-g correspond to the Mumax$_3$ simulation models from Supplementary Materials S5. S-

b: spherical γ-Fe₂O₃ nanoparticle with diameter of 25 nm; S-c: cubic γ-Fe₂O₃ nanoparticle with side length of 15 nm; S-e: the sintered body, spherical γ´-Fe₄N nanoparticle with diameter of 100 nm; S-g: a 5 × 5 array of 100 nm spherical γ´-Fe₄N nanoparticle clustered together. Blue and red color represents +z and –z components of magnetization, respectively. The arrows represent the x-y components of magnetization. The magnetization evolutions of S-a, S-d, and S-f are given in Supplementary Materials S5.

**2.7. Dynamic magnetic responses of γ-Fe2O3 and γ´-Fe4N nanoparticles in aqueous solutions.** The dynamic magnetic responses are characterized by a homebuilt magnetic particle spectroscopy (MPS).[10,26,46–48,48–52] Where the γ-Fe₂O₃ and γ´-Fe₄N nanoparticles suspended in oleic acid (OA) solution (overall volume of 200 μL, concentration of 67 mg/mL) are characterized and recorded by this MPS system. This MPS system generates an alternating current (AC) magnetic field to excite MNPs, as a result, the magnetic moments of nanoparticles relax to align with the external driving field through the Néel- or Brownian-relaxation dominated process or through the joint Néel-Brownian relaxation process (Supplementary Materials S1).[26,46,53,54] In this work, the magnetic moments of both γ-Fe₂O₃ and γ´-Fe₄N nanoparticles in OA relax along the AC magnetic field through a Néel relaxation-dominated process, theoretical analysis can be found from Supplementary Materials S2. The AC magnetic driving field can be tuned with varying frequencies $f$ from 50 Hz to 2850 Hz and the amplitude of the driving field is set at 170 Oe. The relaxation of magnetic moments of MNPs subjected to AC field is dynamic magnetic responses, this time-varying magnetic moment causes electromotive force (EMF) in a pair of differently wound pick-up coils (Faraday's law of induction). As a result, the dynamic magnetic responses of γ-Fe₂O₃ and γ´-Fe₄N nanoparticles are recorded (Supplementary Materials S1) as real-time voltage. Due to the nonlinear magnetic responses of nanoparticles under driving field $f$, higher odd harmonics at $3f$, $5f$, $7f$, etc., are found from the frequency domain of collected voltage signal.[10,47,49,55] Figure 7(a) – (c) summarize the amplitudes of the 3rd, the 5th and the 7th harmonics recorded at $3f$, $5f$ and $7f$, respectively as we vary the driving field frequency from 50 Hz to 2850 Hz. Under all driving field frequencies, the γ´-Fe₄N@BM sample shows stronger magnetic responses (higher harmonic amplitudes) over γ-Fe₂O₃, which indicates that γ´-Fe₄N nanoparticles show higher magnetic moment per particle compared to γ-Fe₂O₃ (discussed in Supplementary Materials S3). The amplitudes of all the odd harmonics collected from both γ-Fe₂O₃@BM and γ´-Fe₄N@BM samples show a similar trend as we vary the driving field frequency: the harmonic amplitude increases as the driving field frequency $f$ increases, it reaches to a plateau at a critical frequency $f_{crit}$ (marked by stars in Figure 7 and Figure 10), then it slowly decays as we further increase the driving field frequency. As the driving field frequency $f$ increases from 50 Hz to 2850 Hz, the γ-Fe₂O₃ and γ´-Fe₄N nanoparticles go through three different regions (labeled as I, II, and III in Figure 7): in region I ($f$-dominant region), the dynamic magnetic responses are dependent on the driving field frequency $f$, the magnetic responses increase as $f$ increases; in region III ($\phi$-dominant region), the dynamic magnetic responses are dependent on the phase lag $\phi$ between the magnetic moments of nanoparticles and the fast-changing AC field; in region II ($f$-$\phi$ co-led), the transitional stage between regions I and III, both $f$ and $\phi$ impact the dynamic magnetic responses and the harmonic amplitude curves reach to their maxima.

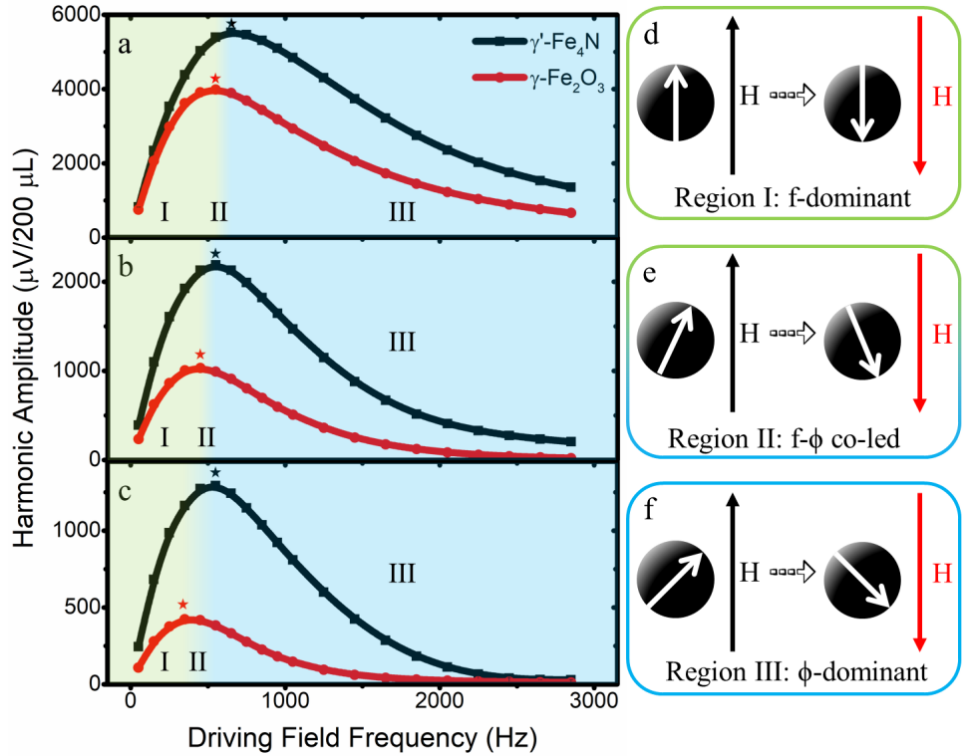

Figure 7. The recorded (a) 3rd, (b) 5th and (c) the 7th harmonic amplitudes as the driving field frequency varies from 50 Hz to 2850 Hz. The red and black stars mark the critical frequencies $f_{crit}$ where γ-Fe2O3 and γ´-Fe4N nanoparticles show highest dynamic magnetic responses. (d) In $f$-dominant region, the magnetic moments of nanoparticles are almost synchronized with the AC magnetic field, the detected harmonic amplitude increases as the driving field frequency $f$ increases. (e) In $f$-$\phi$ co-led region, the dynamic magnetic responses reach to maxima where the enhancement effect of $f$ and the attenuation effect of $\phi$ paly equally important roles. (f) In $\phi$-dominant region, the magnetic moments of nanoparticles cannot synchronize with the fast-changing AC magnetic field, thus a phase lag $\phi$ between the magnetic moments and field causes attenuated harmonic amplitudes detected by the pick-up coils (discussed in Supplementary Materials S3).

**2.8. Real-time dynamic magnetic responses recorded by MPS.** The relaxation of magnetic moments of MNPs subjected to AC driving field causes detectable EMF in the pick-up coils and this EMF is a time-varying voltage signal. This analog voltage signal collected from pick-up coils is sampled at a sampling rate of 500 kHz and the higher odd harmonics (due to the nonlinear dynamic magnetic responses of γ-Fe2O3 and γ´-Fe4N nanoparticles) are extracted. The discrete-time total voltage signal, the 3rd, 5th and the 7th harmonics are replotted in Figure 8(a) – (l). Figure 8(a) – (f) and Figure 8(g) – (l) correspond to the dynamic magnetic responses of γ´-Fe4N and γ-Fe2O3 nanoparticles, respectively. Each time window records the voltage signals within one period of AC driving field (i.e., $1/f$ second). The first to the sixth rows correspond to the scenarios where driving field frequency $f$ =350 Hz, 650 Hz, 950 Hz, 1250 Hz, 1850 Hz, and 2450 Hz, respectively. Distortions in the voltage signal are observed from both γ-Fe$_2$O$_3$@BM and γ´-Fe$_4$N@BM samples (highlighted in grey in Figure 8), where the voltage signal from γ´-Fe$_4$N@BM sample shows severer distortions over γ-Fe$_2$O$_3$@BM sample under the same driving field condition. These distortions are caused by the periodically synchronized higher odd harmonics (the $3f$, $5f$ and $7f$ harmonic voltage signals denoted in red, green, and blue solid curves). Whenever the crests and troughs of higher odd harmonics are synchronized, the cumulative effect causes small convex and concave in the total signal curve, respectively, as highlighted in Figure 8(a) – (l).

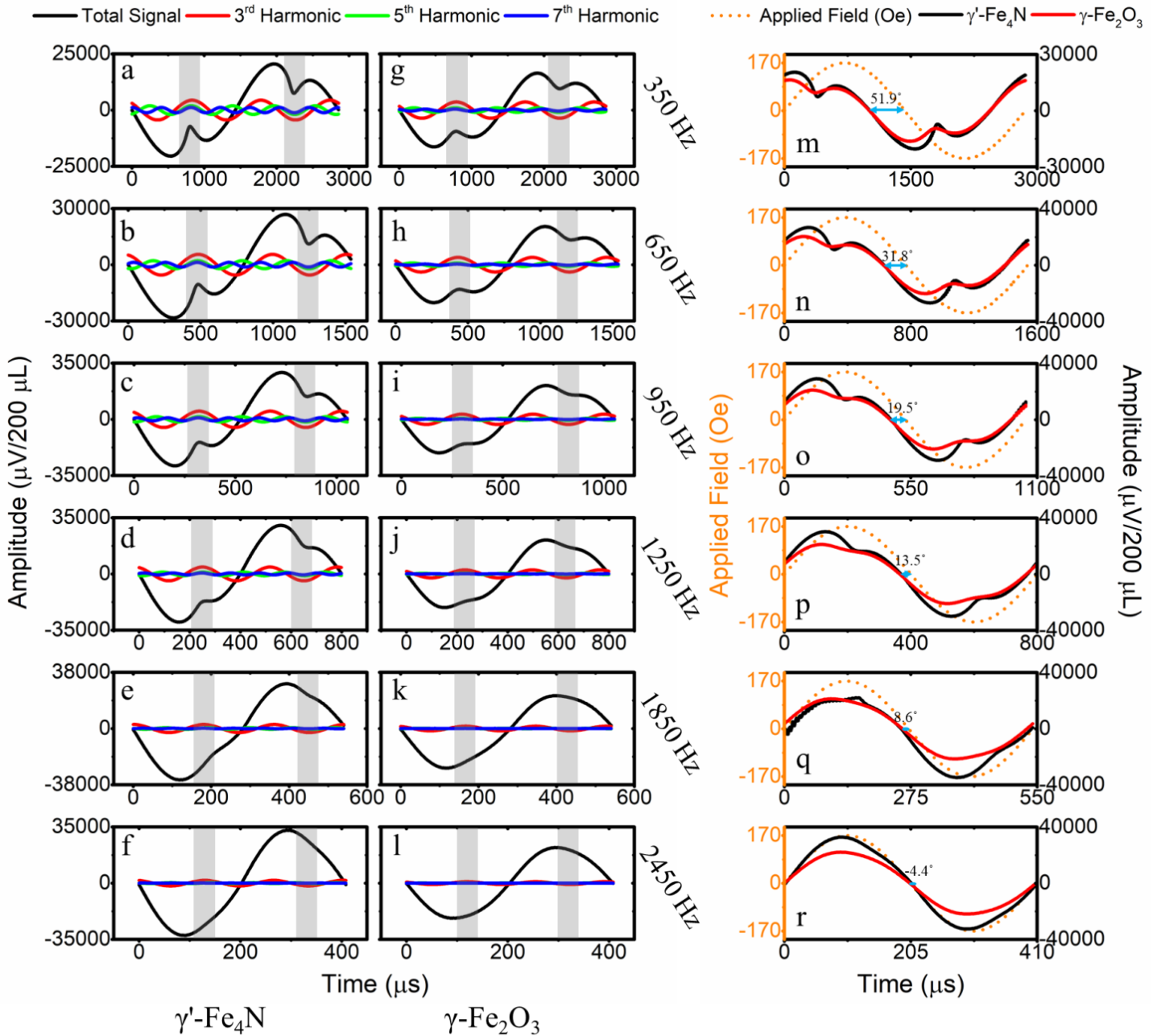

Figure 8. (a) – (l). Real-time magnetic responses recorded by MPS system. The total voltage signal (black solid lines) received a pair of pick-up coils, the 3rd (red solid lines), the 5th (green solid lines) and the 7th (blue solid

lines) harmonics are plotted in a time window of one period of AC driving field ($1/f$ second). (a) – (f) and (g) – (l) are the summarized dynamic magnetic responses from 200 μL, 67 mg/mL γ′-Fe4N@BM and γ-Fe2O3@BM samples, respectively. The distortions in total voltage signal curves (highlighted in grey) are caused by the periodically synchronized higher odd harmonics ($3f$, $5f$ and $7f$). (m) – (r). Total voltage signal from γ-Fe2O3@BM (red solid lines) and γ′-Fe4N@BM (black solid lines) samples plotted along with the AC driving field (orange dotted lines) in time domain. The phase differences between the voltage and AC field under different driving field frequencies are marked by the blue arrows.

**2.9. Phase lags between dynamic magnetic responses and AC magnetic fields.** According to Faraday's law of induction, the time-varying magnetic flux induces EMF in the pick-up coils. During one relaxation process of magnetic moments of nanoparticles to align with the direction of external AC field, the stray field causes the EMF in the coils and there is a 90° phase shift between magnetic moment and detected voltage from pick-up coils. As shown in Figure 8(m) – (r), the real-time voltages collected from γ-Fe2O3@BM and γ′-Fe4N@BM samples along with the AC fields are plotted during a time window of one period of AC driving field. On top of the 90° phase shift due to the law of induction, we observed phase differences of 51.9°, 31.8°, 19.5°, 13.5°, 8.6°, and -4.4° between the voltage and field under the driving field frequencies of 350 Hz, 650 Hz, 950 Hz, 1250 Hz, 1850 Hz, and 2450 Hz, respectively. The detected voltages from γ-Fe2O3@BM and γ′-Fe4N@BM samples are quite synchronous, indicating the identical phase lags of both types of nanoparticles to the AC driving fields. The calculated phase lags of magnetic moments of MNPs to different driving field frequencies are summarized in Figure 9(e).

**2.10. Dynamic magnetization curves of γ-Fe2O3 and γ′-Fe4N nanoparticles.** The dynamic magnetization responses of γ-Fe2O3@BM and γ′-Fe4N@BM samples are calculated using the real-time voltage signals from Figure 8(m) – (r). Figure 9(a) & (b) show the normalized magnetization curves of γ-Fe2O3@BM and γ′-Fe4N@BM samples subjected to driving field frequencies of 350 Hz and 2450 Hz, respectively. At both low and high driving field frequencies, the γ′-Fe4N nanoparticles show higher dynamic magnetic responses than γ-Fe2O3. In addition, as we gradually increase the frequency of driving field, the dynamic hysteresis loops transform from long ellipses to flat ovals for both samples, as shown in Figure 9(c) & (d). This is due to that, as the AC field sweeps faster, both types of nanoparticles are unable to synchronize with the fast-changing AC fields, thus a larger phase lag of magnetic moment to external driving field is induced.

For magnetic hyperthermia treatments, when MNPs are subjected to the AC field, the area of their magnetic hysteresis loop, $A$, corresponds to the dissipated energy.[56–58] The power generated by these MNPs, or specific absorption rate (SAR), is evaluated by the equation, $SAR = A \cdot f$. Since the maximum SAR achievable is directly proportional to the saturation magnetization of MNPs, thus, γ′-Fe4N nanoparticles reported in this work can enhance the SAR and, meanwhile minimize the dose. As shown in Figure 9(a) & (b), the dynamic magnetization curves of γ´-Fe₄N@BM and γ-Fe₂O₃@BM are compared under different driving field frequencies, $f$. The magnetizations are normalized to the magnetizations of γ´-Fe₄N@BM. Under both driving field conditions, γ´-Fe₄N@BM shows larger hysteresis loop area $A$ over γ-Fe₂O₃@BM. Indicating that γ´-Fe₄N@BM could be potentially applied as high-performance heating sources in hyperthermia treatment with minimum dose.

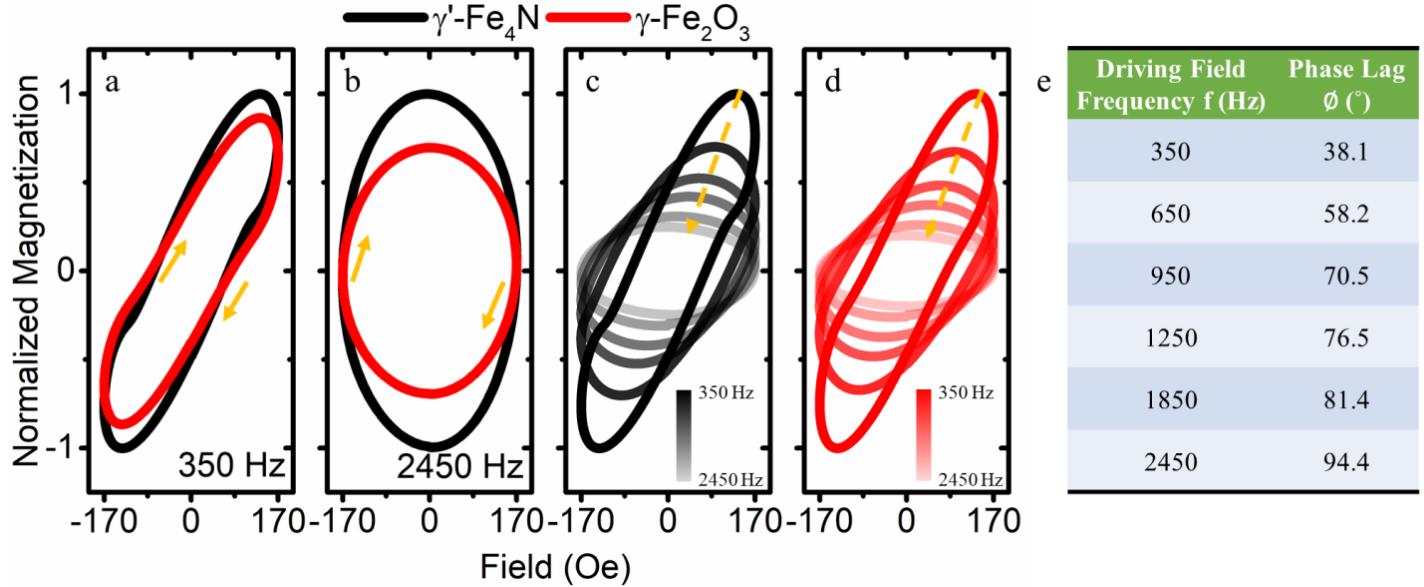

Figure 9. Measured field-dependent dynamic (AC) magnetization curves of γ-Fe₂O₃ and γ´-Fe₄N nanoparticles in OA subjected to (a) 350 Hz and (b) 2450 Hz driving fields. (c) and (d) show the transformations of dynamic magnetization curves as the driving field frequency increases from 350 Hz to 2450 Hz, for γ´-Fe₄N@BM and γ-Fe₂O₃@BM samples, respectively. (e) The calculated phase lags of both MNPs to different driving field frequencies.

**2.11. Normalized harmonics and harmonic ratios.** Since the harmonic amplitudes resulted from the dynamic magnetic responses of nanoparticles are dependent on the quantity of nanoparticles from the testing sample as well as the pick-up coil design (winding number, width and diameter), as discussed in Supplementary Materials S3.[10,47] Thus, the normalized magnetic responses and the harmonic ratios are used as nanoparticle quantity-independent metrics for characterizing the dynamic magnetic properties of nanoparticles.[26,49] Figure 10(a) – (c) show the normalized 3rd, 5th and 7th harmonics from γ´-Fe₄N@BM and γ-Fe₂O₃@BM samples under varying driving field frequencies, corresponding to the recorded harmonic amplitudes in Figure 7(a) – (c). The normalized harmonic curve of γ-Fe₂O₃@BM shows a sharper peak compared to γ´-Fe₄N@BM. The harmonic ratios are summarized in Figure 10(d) – (k) under driving field frequencies of 150 Hz, 350 Hz, 650 Hz, 850 Hz, 1250 Hz, 1650 Hz, 2250 Hz and 2650 Hz. The harmonics of γ´-Fe₄N@BM sample decays at a slower rate as the harmonic number increases (black lines in Figure 10(d) – (k)).

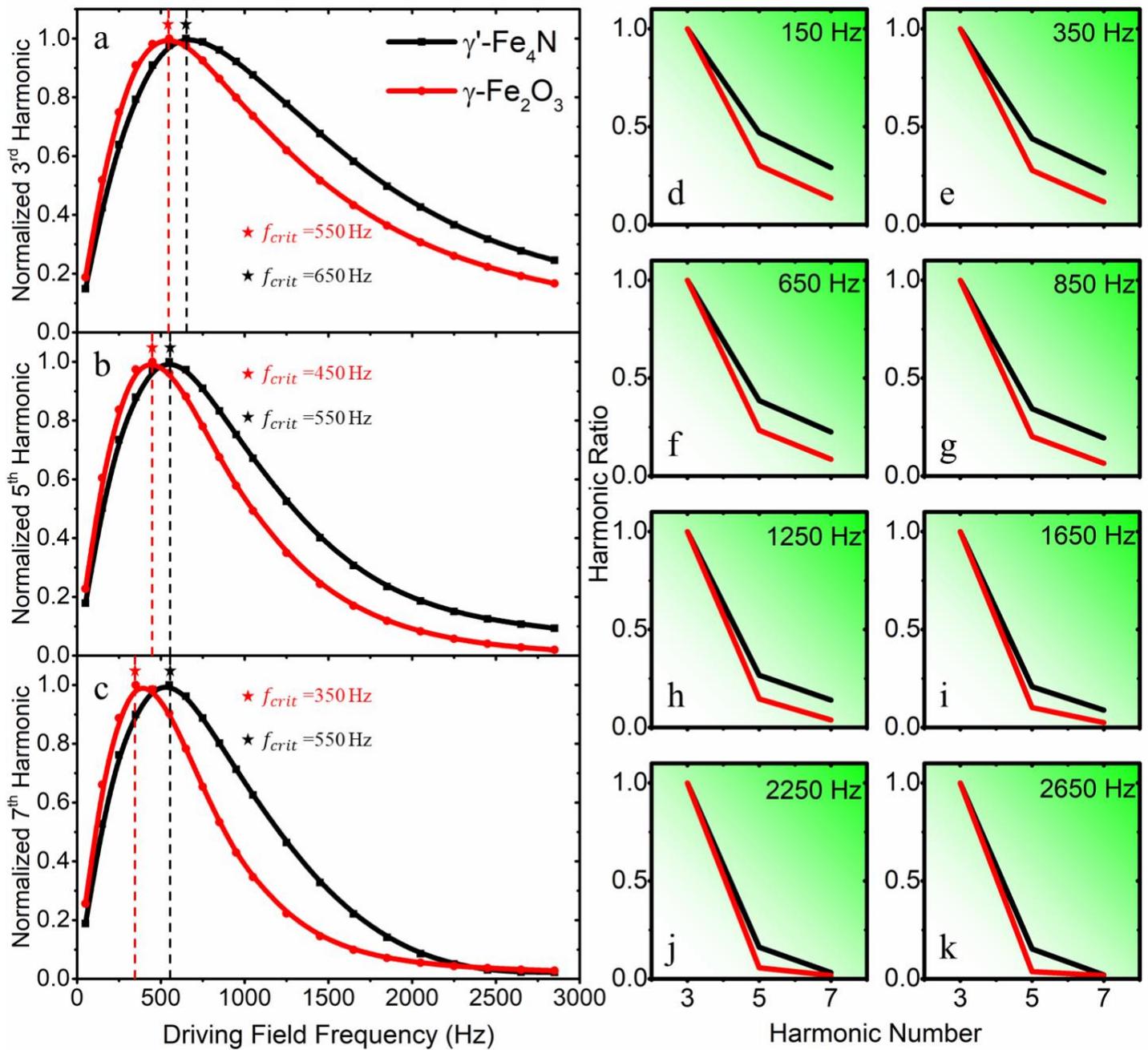

Figure 10. (a) – (c) are the normalized 3rd, 5th and 7th harmonics from γ´-Fe4N and γ-Fe2O3 nanoparticles at varying driving field frequencies, respectively. The critical frequencies $f_{crit}$ at which the harmonics reach to maxima are labeled by stars. (d) – (k) are the harmonic amplitude ratios calculated from γ´-Fe4N and γ-Fe2O3 nanoparticles at varying driving field frequencies.

## 3. CONCLUSIONS

In this paper, we reported high magnetic moment, irregularly shaped γ´-Fe4N nanoparticles with OA surfactant. The crystalline structure, static (DC) and dynamic (AC) magnetic properties are characterized by VSM and MPS systems and compared to the γ-Fe$_2$O$_3$ nanoparticles. Our γ´-Fe$_4$N nanoparticles show superior magnetic properties with more than 3 times higher saturation magnetization when compared to γ-Fe$_2$O$_3$. The dynamic magnetic responses of both nanoparticles are compared and the dynamic (AC) magnetization curves show that our γ´-Fe$_4$N nanoparticles, with large hysteresis loop, could be potentially applied as high-performance heating sources in hyperthermia treatment with minimum dose. In addition, due to the sintering in the nitradation process, the fabricated γ´-Fe$_4$N nanoparticles, with irregular shapes, hold great promise for enhance T$_2$ relaxivity as contrast agents in MRI applications. Since nanoparticles with a larger hydrodynamic size distribution generate greater magnetic field gradient and this leads to a higher order of proton dephasing.[59] The γ´-Fe$_4$N nanoparticles synthesized and characterized in this paper are irregularly shaped and have formed sintered bodies. This will cause them to yield both magnetic fields coupling induced inhomogeneous magnetic field distribution as well as artificially enhanced magnetic field inhomogeneity. This inhomogeneity is extremely advantageous for MRI applications as they pave the way for varied relaxation rates (R$_1$ & R$_2$) and hence varied relaxation times (R$_1$ = 1/T$_1$ & R$_2$ = 1/T$_2$) over a specific area of tissue to be imaged. This phenomenon enhances the contrast efficiency between adjacent tissues for MRI applications.

The stability of OA coated γ-Fe$_2$O$_3$ and γ´-Fe$_4$N nanoparticles are investigated in <span style="color:red">Supplementary Materials S7</span>. With smaller sizes and less aggregations, γ-Fe$_2$O$_3$ nanoparticles show better stability. In this work, OA is used as surfactant on γ´-Fe$_4$N nanoparticles, these lipophilic MNPs show very good dissolvability in polar liquids such as oil. In addition, it is reported that OA can form a dense protective monolayer that binds firmly to the MNP surfaces with enhanced colloidal stability.[60–62] However, for biomedical applications, the lipophilic substances (i.e., OA) coated MNPs are not good candidates and thus the practical use of these MNPs are limited. In the future we can further functionalize OA-coated MNPs with trialkoxysilanes. Thus, the functionalized nanoparticles can be dispersed in various aqueous solutions such as human serum and plasma.[63] Besides, the biocompatibility and colloidal stability of γ´-Fe$_4$N nanoparticles can be further enhanced by conjugating chemical compounds such as chitosan, polyethylene glycol (PEG), amino acids, citric acid, etc., so that the water solubility of MNPs can increase significantly .[64–67]

## 4. EXPERIMENTAL SECTION

**Chemicals.** γ-Fe$_2$O$_3$ powder (purity 99.5%, size 20 nm) is purchase from MTI Corporation, Richmond, CA. Oleic acid is purchased from Fisher Scientific, Hampton, NH. Isopar G fluid is purchased from ExxoMobil, Irving, TX.

**Wet mechanical milling.** 400 mg of γ-Fe$_2$O$_3$ and γ´-Fe$_4$N powders are prepared for wet mechanical milling process, respectively. In the case of wet milling, 400 mg powder is dispersed in 6 mL oleic acid (OA, CH$_3$(CH$_2$)$_7$CH=CH(CH$_2$)$_7$COOH) under argon atmosphere in glove box. The milling conditions are: 14 mm ball diameter, vial rotation of 8000 rpm for 8 hours.

**X-ray diffraction (XRD) characterization.** Samples for the XRD measurement are prepared in the glove box. Certain amount of γ-$Fe_2O_3$ and γ´-$Fe_4N$ nanoparticles are put on a piece of glass and sealed by epoxy to avoid oxidation when the samples are taken out of the glove box for the XRD measurement. XRD pattern are measured using a Bruker D8 discover 2D diffractometer (40 kV and 35 mA). A cobalt radiation source (wavelength~1.79 Å) is used to get better signal. The XRD patterns are converted to copper radiation for a convenient comparison.

**Sample preparation and transmission electron microscopy (TEM) characterization.** The as-milled γ-$Fe_2O_3$ and γ´-$Fe_4N$ nanoparticles in OA (labeled as γ-$Fe_2O_3$@BM and γ´-$Fe_4N$@BM in this paper, with concentration of 67 mg/mL) are ultra-sonicated for 1 hour to make the nanoparticles evenly dispersed. The γ-$Fe_2O_3$@BM and γ´-$Fe_4N$@BM suspensions are diluted by 100 times in OA, then 10 μL of each suspension (concentration of 0.67 mg/mL) is dropped onto TEM grids. The γ-$Fe_2O_3$@BM and γ´-$Fe_4N$@BM in OA are ultra-centrifuged at 10,000 rpm, 9,300 g for 20 minutes, then 10 μL supernatant (labeled as γ-$Fe_2O_3$@Ultra and γ´-$Fe_4N$@Ultra in this paper) is drew from each sample and dropped onto TEM grids. The γ-$Fe_2O_3$@BM and γ´-$Fe_4N$@BM in OA are sealed and placed at room temperature for 24 hours, the larger MNP clusters precipitate to the bottom of suspension and 10 μL supernatant (labeled as γ-$Fe_2O_3$@Sup and γ´-$Fe_4N$@Sup in this paper) is drew from each sample and dropped onto TEM grids. All the 6 TEM grids are air dried before the TEM characterizations. A FEI Tecnai T12 transmission electron microscopy (T12, 120 kV) is used to characterize the samples.

**Dynamic light scattering (DLS) and nanoparticle tracking analyzer (NTA) characterization.** 50 μL of wet mechanically milled γ-$Fe_2O_3$ (transparent, refractive index, r.i = 2.91) and γ´-$Fe_4N$ (absorbing, irregularly shaped) in OA are diluted by 8 times in a synthetic isoparaffinic fluid, Isopar G fluid (r.i = 1.49). The hydrodynamic size distribution of the both γ-$Fe_2O_3$ and γ´-$Fe_4N$ nanoparticles are characterized by the DLS particle analyzer (Model name: Microtrac NanoFlex).

To establish a stronger corroboration with that of the TEM images and DLS results, the samples of the same order of dilution were characterized in nanoparticle tracking analyzer (NTA, Model: Nanosight LM-10). We would like to clarify here that the hydrodynamic size distribution obtained from DLS for both γ-$Fe_2O_3$ and γ´-$Fe_4N$ are in correlation with the TEM images. However, only γ´-$Fe_4N$ could be characterized by NTA. Although the reason behind why γ-$Fe_2O_3$ could not be characterized by NTA is not clear, our estimation is that as NTA uses a wavelength of 403 nm, γ-$Fe_2O_3$ having an extremely high transparent refractive index (r.i. = 2.91) in Isorpar G fluid media of much lower refractive index must cause total internal reflection for most of the light. Nevertheless, NTA gives us the concentration of the γ´-$Fe_4N$ nanoparticles which in turn specifies the γ-$Fe_2O_3$ concentration as both the nanoparticle samples for hydrodynamic size characterization, were prepared in a similar manner. The sample volume prepared for both DLS as well as NTA particle analyzer is 1.5 mL.

**Vibrating sample magnetometer (VSM) characterization.** 25 μL γ-$Fe_2O_3$ and γ´-$Fe_4N$ nanoparticles in OA suspensions are dropped on a filter paper and air-dried, then fit into a 5 mm diameter 12 mm long gelatin capsule (gelcap). During the VSM measurement, the gelcap is inserted into sample tube and affixed to the sample-rod. The magnetic field is swept from -5000 to +5000 Oe with a step width of 5 Oe (or -2000 to +2000 Oe with a step width of 2 Oe), the averaging time for each measurement is 100 ms.

**Mumax3 Simulation.** The TEM images, DLS & NTA characterizations give us a legitimate knowledge on the idea about the shapes and sizes of γ-$Fe_2O_3$ and fabricated γ´-$Fe_4N$ nanoparticles. We simulate the shapes of the nanoparticles using micromagnetic framework, Mumax3, and observe the magnetization distribution within the γ-$Fe_2O_3$ and γ´-$Fe_4N$ nanoparticles.[68] The magnetic properties of the nanoparticles were obtained from our experimental and previously reported results listed in Table S1 and Table S2 in the Supplementary Materials S4. A total of seven different nanoparticle shapes are modeled and plotted in Supplementary Materials S5. Four different shapes of γ-$Fe_2O_3$ nanoparticles are modeled: spherical with diameter of 15 nm (labeled as S-a), spherical with diameter of 25 nm (labeled as S-b), cubic with side length of 15 nm (labeled as S-c), and ellipsoid with long axis of 30 nm and short axis of 10 nm (labeled as S-d). Three different shapes of γ´-$Fe_4N$ nanoparticles are modeled: sintered body, spherical with diameter of 100 nm (labeled as S-e), ellipsoid with long axis of 200 nm and short axis of 50 nm (labeled as S-f), a 5 × 5 array of 100 nm spherical γ´-$Fe_4N$ nanoparticles clustered together (labeled as S-g). The uniaxial anisotropy γ-$Fe_2O_3$ nanoparticles with easy axis align along [1 1 1] and cubic anisotropy γ´-$Fe_4N$ nanoparticles with easy axis align along [1 0 0] are assumed, external magnetic field is applied along [1 1 0] direction (see Supplementary Materials S5). The mathematical models for uniaxial and cubic anisotropy energy distributions are given in Supplementary Materials S6.

**Magnetic particle spectroscopy (MPS) measurement.** 200 μL γ-$Fe_2O_3$ and γ´-$Fe_4N$ nanoparticles in OA suspensions (concentration of 67 mg/mL) are sealed in a plastic vial for MPS measurements. The AC magnetic field frequency $f$ is varied from 50 Hz to 2850 Hz, with amplitude set at 170 Oe. For each run, the vial containing nanoparticles is inserted into the pick-up coils (see Supplementary Materials S1) and real time voltage signal is collected for 10 seconds. The analog voltage signal is sampled at a sampling rate of 500 kHz.

**Supporting Information**

Magnetic particle spectroscopy (MPS) system setups and magnetic relaxation mechanisms of magnetic nanoparticles under AC driving fields; Néel relaxation-dominated γ-$Fe_2O_3$@BM and γ´-$Fe_4N$@BM MNPs in oleic acid; Models of dynamic magnetic responses; Micromagnetic simulation parameters; Micromagnetic simulation models; Magnetocrystalline anisotropy energy of γ-$Fe_2O_3$ and γ´-$Fe_4N$; Stability of γ-$Fe_2O_3$ and γ´-$Fe_4N$ nanoparticles in oleic acid.

Supplementary Videos: Figure 5 γ´-$Fe_4N$ NTA curves I – V

# AUTHOR INFORMATION
**Corresponding Authors**

*E-mail: wuxx0803@umn.edu (K.W.).

*E-mail: jpwang@umn.edu (J.-P.W.).

**ORCID**

Kai Wu: 0000-0002-9444-6112

Jinming Liu: 0000-0002-4313-5816

Renata Saha: 0000-0002-0389-0083

Diqing Su: 0000-0002-5790-8744

Chaoyi Peng: 0000-0003-1608-3886

Jian-Ping Wang: 0000-0003-2815-6624

**Author Contributions**

⊥K.W., J.L., and R.S. contributed equally to this work.

**Notes**

The authors declare no conflict of interest.

**Acknowledgments**

This study was financially supported by the Institute of Engineering in Medicine of the University of Minnesota through FY18 IEM Seed Grant Funding Program. Portions of this work were conducted in the Minnesota Nano Center, which is supported by the National Science Foundation through the National Nano Coordinated Infrastructure Network (NNCI) under Award Number ECCS-1542202. Portions of this work were carried out in the Characterization Facility, University of Minnesota, a member of the NSF-funded Materials Research Facilities Network (www.mrfn.org) via the MRSEC program.

**References**

(1) Adolphi, N. L.; Huber, D. L.; Bryant, H. C.; Monson, T. C.; Fegan, D. L.; Lim, J.; Trujillo, J. E.; Tessier, T. E.; Lovato, D. M.; Butler, K. S. Characterization of Single-Core Magnetite Nanoparticles for Magnetic Imaging by SQUID Relaxometry. *Phys. Med. Biol.* **2010**, *55* (19), 5985.

(2) Arami, H.; Khandhar, A. P.; Tomitaka, A.; Yu, E.; Goodwill, P. W.; Conolly, S. M.; Krishnan, K. M. In Vivo Multimodal Magnetic Particle Imaging (MPI) with Tailored Magneto/Optical Contrast Agents. *Biomaterials* **2015**, *52*, 251–261.

(3) Bauer, L. M.; Situ, S. F.; Griswold, M. A.; Samia, A. C. S. Magnetic Particle Imaging Tracers: State-of-the-Art and Future Directions. *J Phys Chem Lett* **2015**, *6* (13), 2509–2517.

(4) Chieh, J.-J.; Huang, K.-W.; Lee, Y.-Y.; Wei, W.-C. Dual-Imaging Model of SQUID Biosusceptometry for Locating Tumors Targeted Using Magnetic Nanoparticles. *J. Nanobiotechnology* **2015**, *13* (1), 1.

(5) Du, Y.; Lai, P. T.; Leung, C. H.; Pong, P. W. Design of Superparamagnetic Nanoparticles for Magnetic Particle Imaging (MPI). *Int. J. Mol. Sci.* **2013**, *14* (9), 18682–18710.

(6) Iqbal, M. Z.; Ma, X.; Chen, T.; Zhang, L.; Ren, W.; Xiang, L.; Wu, A. Silica-Coated Super-Paramagnetic Iron Oxide Nanoparticles (SPIONPs): A New Type Contrast Agent of T 1 Magnetic Resonance Imaging (MRI). *J. Mater. Chem. B* **2015**, *3* (26), 5172–5181.

(7) Mishra, S. K.; Kumar, B.; Khushu, S.; Tripathi, R. P.; Gangenahalli, G. Increased Transverse Relaxivity in Ultrasmall Superparamagnetic Iron Oxide Nanoparticles Used as MRI Contrast Agent for Biomedical Imaging. *Contrast Media Mol. Imaging* **2016**.

(8) Panagiotopoulos, N.; Duschka, R. L.; Ahlborg, M.; Bringout, G.; Debbeler, C.; Graeser, M.; Kaethner, C.; Lüdtke-Buzug, K.; Medimagh, H.; Stelzner, J. Magnetic Particle Imaging: Current Developments and Future Directions. *Int. J. Nanomedicine* **2015**, *10*, 3097.

(9) Shen, Z.; Wu, A.; Chen, X. Iron Oxide Nanoparticle Based Contrast Agents for Magnetic Resonance Imaging. *Mol. Pharm.* **2016**, *14* (5), 1352–1364.

(10) Wu, K.; Su, D.; Saha, R.; Wong, D.; Wang, J.-P. Magnetic Particle Spectroscopy-Based Bioassays: Methods, Applications, Advances, and Future Opportunities. *J. Phys. Appl. Phys.* **2019**, *52*, 173001.

(11) Arruebo, M.; Fernández-Pacheco, R.; Ibarra, M. R.; Santamaría, J. Magnetic Nanoparticles for Drug Delivery. *Nano Today* **6**, *2* (3), 22–32. https://doi.org/10.1016/S1748-0132(07)70084-1.

(12) Chen, Y.; Ai, K.; Liu, J.; Sun, G.; Yin, Q.; Lu, L. Multifunctional Envelope-Type Mesoporous Silica Nanoparticles for PH-Responsive Drug Delivery and Magnetic Resonance Imaging. *Biomaterials* **2015**, *60*, 111–120.

(13) Chowdhuri, A. R.; Bhattacharya, D.; Sahu, S. K. Magnetic Nanoscale Metal Organic Frameworks for Potential Targeted Anticancer Drug Delivery, Imaging and as an MRI Contrast Agent. *Dalton Trans.* **2016**, *45* (7), 2963–2973.

(14) Ding, Y.; Shen, S. Z.; Sun, H.; Sun, K.; Liu, F.; Qi, Y.; Yan, J. Design and Construction of Polymerized-Chitosan Coated Fe3O4 Magnetic Nanoparticles and Its Application for Hydrophobic Drug Delivery. *Mater. Sci. Eng. C* **2015**, *48*, 487–498.

(15) Estelrich, J.; Escribano, E.; Queralt, J.; Busquets, M. A. Iron Oxide Nanoparticles for Magnetically-Guided and Magnetically-Responsive Drug Delivery. *Int. J. Mol. Sci.* **2015**, *16* (4), 8070–8101.

(16) Bao, G.; Mitragotri, S.; Tong, S. Multifunctional Nanoparticles for Drug Delivery and Molecular Imaging. *Annu. Rev. Biomed. Eng.* **2013**, *15*, 253–282.

(17) Yu, L.; Liu, J.; Wu, K.; Klein, T.; Jiang, Y.; Wang, J.-P. Evaluation of Hyperthermia of Magnetic Nanoparticles by Dehydrating DNA. *Sci. Rep.* **2014**, *4*, 7216.

(18) Andreu, I.; Natividad, E.; Solozábal, L.; Roubeau, O. Nano-Objects for Addressing the Control of Nanoparticle Arrangement and Performance in Magnetic Hyperthermia. *ACS Nano* **2015**, *9* (2), 1408–1419.

(19) Bañobre-López, M.; Teijeiro, A.; Rivas, J. Magnetic Nanoparticle-Based Hyperthermia for Cancer Treatment. *Rep. Pract. Oncol. Radiother.* **2013**, *18* (6), 397–400.

(20) Blanco-Andujar, C.; Teran, F.; Ortega, D. Current Outlook and Perspectives on Nanoparticle-Mediated Magnetic Hyperthermia. In *Iron Oxide Nanoparticles for Biomedical Applications*; Elsevier, 2018; pp 197–245.

(21) Carrey, J.; Mehdaoui, B.; Respaud, M. Simple Models for Dynamic Hysteresis Loop Calculations of Magnetic Single-Domain Nanoparticles: Application to Magnetic Hyperthermia Optimization. *J. Appl. Phys.* **2011**, *109* (8), 083921.

(22) Coral, D. F.; Mendoza Zélis, P.; Marciello, M.; Morales, M. D. P.; Craievich, A.; Sanchez, F. H.; Fernández van Raap, M. B. On the Effect of Nanoclustering and Dipolar Interactions in Heat Generation for Magnetic Hyperthermia. *Langmuir* **2016**.

(23) Chen, S.; Chiang, C.; Hsieh, S. Simulating Physiological Conditions to Evaluate Nanoparticles for Magnetic Fluid Hyperthermia (MFH) Therapy Applications. *J. Magn. Magn. Mater.* **2010**, *322*, 247–252.

(24) Murase, K.; Aoki, M.; Banura, N.; Nishimoto, K.; Mimura, A.; Kuboyabu, T.; Yabata, I. Usefulness of Magnetic Particle Imaging for Predicting the Therapeutic Effect of Magnetic Hyperthermia. *Open J. Med. Imaging* **2015**, *5* (02), 85.

(25) Wu, K.; Wang, J.-P. Magnetic Hyperthermia Performance of Magnetite Nanoparticle Assemblies under Different Driving Fields. *AIP Adv.* **2017**, *7* (5), 056327.

(26) Wu, K.; Liu, J.; Su, D.; Saha, R.; Wang, J.-P. Magnetic Nanoparticle Relaxation Dynamics-Based Magnetic Particle Spectroscopy for Rapid and Wash-Free Molecular Sensing. *ACS Appl. Mater. Interfaces* **2019**, *11* (26), 22979–22986. https://doi.org/10.1021/acsami.9b05233.

(27) Wu, K.; Klein, T.; Krishna, V. D.; Su, D.; Perez, A. M.; Wang, J.-P. Portable GMR Handheld Platform for the Detection of Influenza A Virus. *ACS Sens.* **2017**, *2*, 1594–1601.

(28) Choi, J.; Gani, A. W.; Bechstein, D. J.; Lee, J.-R.; Utz, P. J.; Wang, S. X. Portable, One-Step, and Rapid GMR Biosensor Platform with Smartphone Interface. *Biosens. Bioelectron.* **2016**, *85*, 1–7.

(29) Klein, T.; Wang, Y.; Tu, L.; Yu, L.; Feng, Y.; Wang, W.; Wang, J.-P. Comparative Analysis of Several GMR Strip Sensor Configurations for Biological Applications. *Sens. Actuators Phys.* **2014**, *216*, 349–354.

(30) Tian, B.; De La Torre, T. Z. G.; Donolato, M.; Hansen, M. F.; Svedlindh, P.; Strömberg, M. Multi-Scale Magnetic Nanoparticle Based Optomagnetic Bioassay for Sensitive DNA and Bacteria Detection. *Anal. Methods* **2016**, *8*, 5009–5016.

(31) Chen, Y.; Xianyu, Y.; Wang, Y.; Zhang, X.; Cha, R.; Sun, J.; Jiang, X. One-Step Detection of Pathogens and Viruses: Combining Magnetic Relaxation Switching and Magnetic Separation. *ACS Nano* **2015**, *9* (3), 3184–3191.

(32) Herrmann, I.; Schlegel, A.; Graf, R.; Stark, W. J.; Beck-Schimmer, B. Magnetic Separation-Based Blood Purification: A Promising New Approach for the Removal of Disease-Causing Compounds? *J. Nanobiotechnology* **2015**, *13* (1), 49.

(33) Iranmanesh, M.; Hulliger, J. Magnetic Separation: Its Application in Mining, Waste Purification, Medicine, Biochemistry and Chemistry. *Chem. Soc. Rev.* **2017**, *46* (19), 5925–5934.

(34) Zhang, X.; Wang, J.; Li, R.; Dai, Q.; Gao, R.; Liu, Q.; Zhang, M. Preparation of Fe3O4@ C@ Layered Double Hydroxide Composite for Magnetic Separation of Uranium. *Ind. Eng. Chem. Res.* **2013**, *52* (30), 10152–10159.

(35) Inglis, D. W.; Riehn, R.; Austin, R.; Sturm, J. Continuous Microfluidic Immunomagnetic Cell Separation. *Appl. Phys. Lett.* **2004**, *85* (21), 5093–5095.

(36) Tagawa, K.; Kita, E.; Tasaki, A. Synthesis of Fine Fe4N Powder and Its Magnetic Characteristics. *Jpn. J. Appl. Phys.* **1982**, *21* (11R), 1596.

(37) Wu, X.; Zhong, W.; Jiang, H.; Tang, N.; Zou, W.; Du, Y. Magnetic Properties and Thermal Stability of Γ′-Fe4N Nanoparticles Prepared by a Combined Method of Reduction and Nitriding. *J. Magn. Magn. Mater.* **2004**, *281* (1), 77–81.

(38) Dhanasekaran, P.; Salunke, H. G.; Gupta, N. M. Visible-Light-Induced Photosplitting of Water over Γ′-Fe4N and Γ′-Fe4N/α-Fe2O3 Nanocatalysts. *J. Phys. Chem. C* **2012**, *116* (22), 12156–12164.

(39) Yu, M.; Xu, Y.; Mao, Q.; Li, F.; Wang, C. Electromagnetic and Absorption Properties of Nano-Sized and Micro-Sized Fe4N Particles. *J. Alloys Compd.* **2016**, *656*, 362–367.

(40) Van Voorthuysen, E. D. M.; Boerma, D.; Chechenin, N. Low-Temperature Extension of the Lehrer Diagram and the Iron-Nitrogen Phase Diagram. *Metall. Mater. Trans. A* **2002**, *33* (8), 2593–2598.

(41) Kooi, B. J.; Somers, M. A.; Mittemeijer, E. J. An Evaluation of the Fe-N Phase Diagram Considering Long-Range Order of N Atoms in Γ′-Fe 4 N 1-x and ε-Fe 2 N 1-z. *Metall. Mater. Trans. A* **1996**, *27* (4), 1063–1071.

(42) Liu, J.; Guo, G.; Zhang, F.; Wu, Y.; Ma, B.; Wang, J.-P. Synthesis of A″-Fe 16 N 2 Ribbons with a Porous Structure. *Nanoscale Adv.* **2019**, *1* (4), 1337–1342.

(43) Yamaguchi, T.; Sakita, M.; Nakamura, M.; Kobira, T. Synthesis and Characterics of Fe4N Powders and Thin Films. *J. Magn. Magn. Mater.* **2000**, *215*, 529–531.

(44) Huang, M.; Wallace, W.; Simizu, S.; Pedziwiatr, A.; Obermyer, R.; Sankar, S. Synthesis and Characterization of Fe16N2 in Bulk Form. *J. Appl. Phys.* **1994**, *75* (10), 6574–6576.

(45) Yu, Z. Q.; Zhang, J. R.; Du, Y. W. Magnetic Properties and Preparation of Fe3N Compound. *J. Magn. Magn. Mater.* **1996**, *159* (1–2), L8–L10.

(46) Tu, L.; Wu, K.; Klein, T.; Wang, J.-P. Magnetic Nanoparticles Colourization by a Mixing-Frequency Method. *J. Phys. Appl. Phys.* **2014**, *47*, 155001.

(47) Krause, H.-J.; Wolters, N.; Zhang, Y.; Offenhäusser, A.; Miethe, P.; Meyer, M. H.; Hartmann, M.; Keusgen, M. Magnetic Particle Detection by Frequency Mixing for Immunoassay Applications. *J. Magn. Magn. Mater.* **2007**, *311*, 436–444.

(48) Achtsnicht, S.; Pourshahidi, A. M.; Offenhäusser, A.; Krause, H.-J. Multiplex Detection of Different Magnetic Beads Using Frequency Scanning in Magnetic Frequency Mixing Technique. *Sensors* **2019**, *19* (11), 2599.

(49) Zhang, X.; Reeves, D. B.; Perreard, I. M.; Kett, W. C.; Griswold, K. E.; Gimi, B.; Weaver, J. B. Molecular Sensing with Magnetic Nanoparticles Using Magnetic Spectroscopy of Nanoparticle Brownian Motion. *Biosens. Bioelectron.* **2013**, *50*, 441–446.

(50) Khurshid, H.; Shi, Y.; Berwin, B. L.; Weaver, J. B. Evaluating Blood Clot Progression Using Magnetic Particle Spectroscopy. *Med. Phys.* **2018**, *45* (7), 3258–3263.

(51) Viereck, T.; Draack, S.; Schilling, M.; Ludwig, F. Multi-Spectral Magnetic Particle Spectroscopy for the Investigation of Particle Mixtures. *J. Magn. Magn. Mater.* **2019**, *475*, 647–651.

(52) Draack, S.; Lucht, N.; Remmer, H.; Martens, M.; Fischer, B.; Schilling, M.; Ludwig, F.; Viereck, T. Multiparametric Magnetic Particle Spectroscopy of CoFe2O4 Nanoparticles in Viscous Media. *J. Phys. Chem. C* **2019**, *123* (11), 6787–6801.

(53) Wu, K.; Liu, J.; Wang, Y.; Ye, C.; Feng, Y.; Wang, J.-P. Superparamagnetic Nanoparticle-Based Viscosity Test. *Appl. Phys. Lett.* **2015**, *107*, 053701.

(54) Wu, K.; Schliep, K.; Zhang, X.; Liu, J.; Ma, B.; Wang, J. Characterizing Physical Properties of Superparamagnetic Nanoparticles in Liquid Phase Using Brownian Relaxation. *Small* **2017**, *13*, 1604135.

(55) Nikitin, P. I.; Vetoshko, P. M.; Ksenevich, T. I. New Type of Biosensor Based on Magnetic Nanoparticle Detection. *J. Magn. Magn. Mater.* **2007**, *311*, 445–449.

(56) Mehdaoui, B.; Meffre, A.; Carrey, J.; Lachaize, S.; Lacroix, L.; Gougeon, M.; Chaudret, B.; Respaud, M. Optimal Size of Nanoparticles for Magnetic Hyperthermia: A Combined Theoretical and Experimental Study. *Adv. Funct. Mater.* **2011**, *21* (23), 4573–4581.

(57) He, S.; Zhang, H.; Liu, Y.; Sun, F.; Yu, X.; Li, X.; Zhang, L.; Wang, L.; Mao, K.; Wang, G. Maximizing Specific Loss Power for Magnetic Hyperthermia by Hard–Soft Mixed Ferrites. *Small* **2018**, *14* (29), 1800135.

(58) Asensio, J. M.; Marbaix, J.; Mille, N.; Lacroix, L.-M.; Soulantica, K.; Fazzini, P.-F.; Carrey, J.; Chaudret, B. To Heat or Not to Heat: A Study of the Performances of Iron Carbide Nanoparticles in Magnetic Heating. *Nanoscale* **2019**, *11* (12), 5402–5411.

(59) Zhou, Z.; Tian, R.; Wang, Z.; Yang, Z.; Liu, Y.; Liu, G.; Wang, R.; Gao, J.; Song, J.; Nie, L. Artificial Local Magnetic Field Inhomogeneity Enhances T2 Relaxivity. *Nat. Commun.* **2017**, *8*.

(60) Yang, K.; Peng, H.; Wen, Y.; Li, N. Re-Examination of Characteristic FTIR Spectrum of Secondary Layer in Bilayer Oleic Acid-Coated Fe3O4 Nanoparticles. *Appl. Surf. Sci.* **2010**, *256* (10), 3093–3097.

(61) Shete, P.; Patil, R.; Tiwale, B.; Pawar, S. Water Dispersible Oleic Acid-Coated Fe3O4 Nanoparticles for Biomedical Applications. *J. Magn. Magn. Mater.* **2015**, *377*, 406–410.

(62) Marinca, T.; Chicinaş, H.; Neamţu, B.; Isnard, O.; Pascuta, P.; Lupu, N.; Stoian, G.; Chicinaş, I. Mechanosynthesis, Structural, Thermal and Magnetic Characteristics of Oleic Acid Coated Fe3O4 Nanoparticles. *Mater. Chem. Phys.* **2016**, *171*, 336–345.

(63) Bloemen, M.; Brullot, W.; Luong, T. T.; Geukens, N.; Gils, A.; Verbiest, T. Improved Functionalization of Oleic Acid-Coated Iron Oxide Nanoparticles for Biomedical Applications. *J. Nanoparticle Res.* **2012**, *14* (9), 1100.

(64) Li, X.; Wei, J.; Aifantis, K. E.; Fan, Y.; Feng, Q.; Cui, F.; Watari, F. Current Investigations into Magnetic Nanoparticles for Biomedical Applications. *J. Biomed. Mater. Res. A* **2016**, *104* (5), 1285–1296.

(65) Jin, Y.; Liu, F.; Shan, C.; Tong, M.; Hou, Y. Efficient Bacterial Capture with Amino Acid Modified Magnetic Nanoparticles. *Water Res.* **2014**, *50*, 124–134.

(66) Lin, J.-F.; Tsai, C.-C.; Lee, M.-Z. Linear Birefringence and Dichroism in Citric Acid Coated Fe3O4 Magnetic Nanoparticles. *J. Magn. Magn. Mater.* **2014**, *372*, 147–158.

(67) Mohapatra, S.; Mallick, S.; Maiti, T.; Ghosh, S.; Pramanik, P. Synthesis of Highly Stable Folic Acid Conjugated Magnetite Nanoparticles for Targeting Cancer Cells. *Nanotechnology* **2007**, *18* (38), 385102.

(68) Vansteenkiste, A.; Leliaert, J.; Dvornik, M.; Helsen, M.; Garcia-Sanchez, F.; Van Waeyenberge, B. The Design and Verification of MuMax3. *AIP Adv.* **2014**, *4* (10), 107133.

TOC:

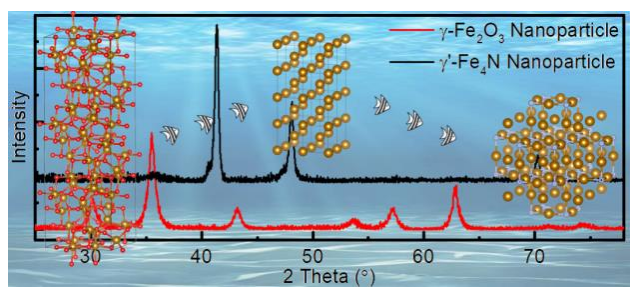

# Supplementary Materials

# Irregularly Shaped γ´-Fe$_4$N Nanoparticles for Hyperthermia Treatment and T$_2$ Contrast-Enhanced Magnetic Resonance Imaging with Minimum Dose

Kai Wu[†,⊥,*], Jinming Liu[†,⊥], Renata Saha[†,⊥], Bin Ma[†], Diqing Su[‡], Chaoyi Peng[†], Jiajia Sun[†,§], and Jian-Ping Wang[†,*]

[†]Department of Electrical and Computer Engineering, University of Minnesota, Minneapolis, Minnesota 55455, USA

[‡]Department of Chemical Engineering and Material Science, University of Minnesota, Minneapolis, Minnesota 55455, USA

[§]State Key Laboratory of Electrical Insulation and Power Equipment, Xi'an Jiaotong University, Xi'an, Shaanxi Province 710049, China

**S1. Magnetic particle spectroscopy (MPS) system setups and magnetic relaxation mechanisms of magnetic nanoparticles under AC driving fields.**

As shown in Figure S1(a), the homebuilt MPS system consists of personal computer (PC), data acquisition card (DAQ, NI USB-6289), instrument amplifier (IA, HP 6824A), one set of copper coil with 1200 windings to generate AC magnetic fields, and one pair of pick-up coils with 600 windings clock-wise and 600 windings counter-clock-wise placed in the center to pick up the magnetic responses of magnetic nanoparticles (MNPs).[1,1–8] A plastic vial containing 200 μL of γ-$Fe_2O_3$@BM or γ′-$Fe_4N$@BM in OA is placed in the upper half part of the pick-up coils. The dynamic magnetic responses of MNPs induces electromotive force (EMF) in the pick-up coil (Faraday's law of induction), this time-varying voltage is sent back to DAQ for the analog to digital conversion (ADC) and Discrete-time Fourier Transform (DTFT) after being filtered. Both the discrete time voltage signal (after ADC) and the frequency domain spectra (after DTFT) are saved for the analysis of the dynamic magnetic responses of MNP samples.

As the AC magnetic field sweeps from -170 Oe to +170 Oe, the magnetic moments of MNPs (either γ-$Fe_2O_3$@BM or γ′-$Fe_4N$@BM MNPs) relax to align with the direction of this driving field, to minimize the magnetostatic energy. This relaxation process can be divided into two different mechanisms: one is the intrinsic Néel relaxation (rotating magnetic moment inside the stationary MNP) and the other is the extrinsic Brownian relaxation (rotating the entire MNP along with its magnetic moment), as shown in Figure S1(b)&(c).

Due to the nonlinear dynamic magnetic responses of MNPs under AC driving fields (see the AC/dynamic magnetization curves in Figure 10 in the paper), higher odd harmonics at $3f$, $5f$, $7f$, etc., are found from the frequency domain of collected responses,[4,9–11] as shown in Figure S1(d)&(e). The intensities (amplitudes)of these higher harmonics provide us with the dynamic magnetic responses and magnetic properties of MNPs under AC driving field, which the DC/static magnetization measurements (Figure 6(a)&(b) in the paper) by VSM cannot provide.

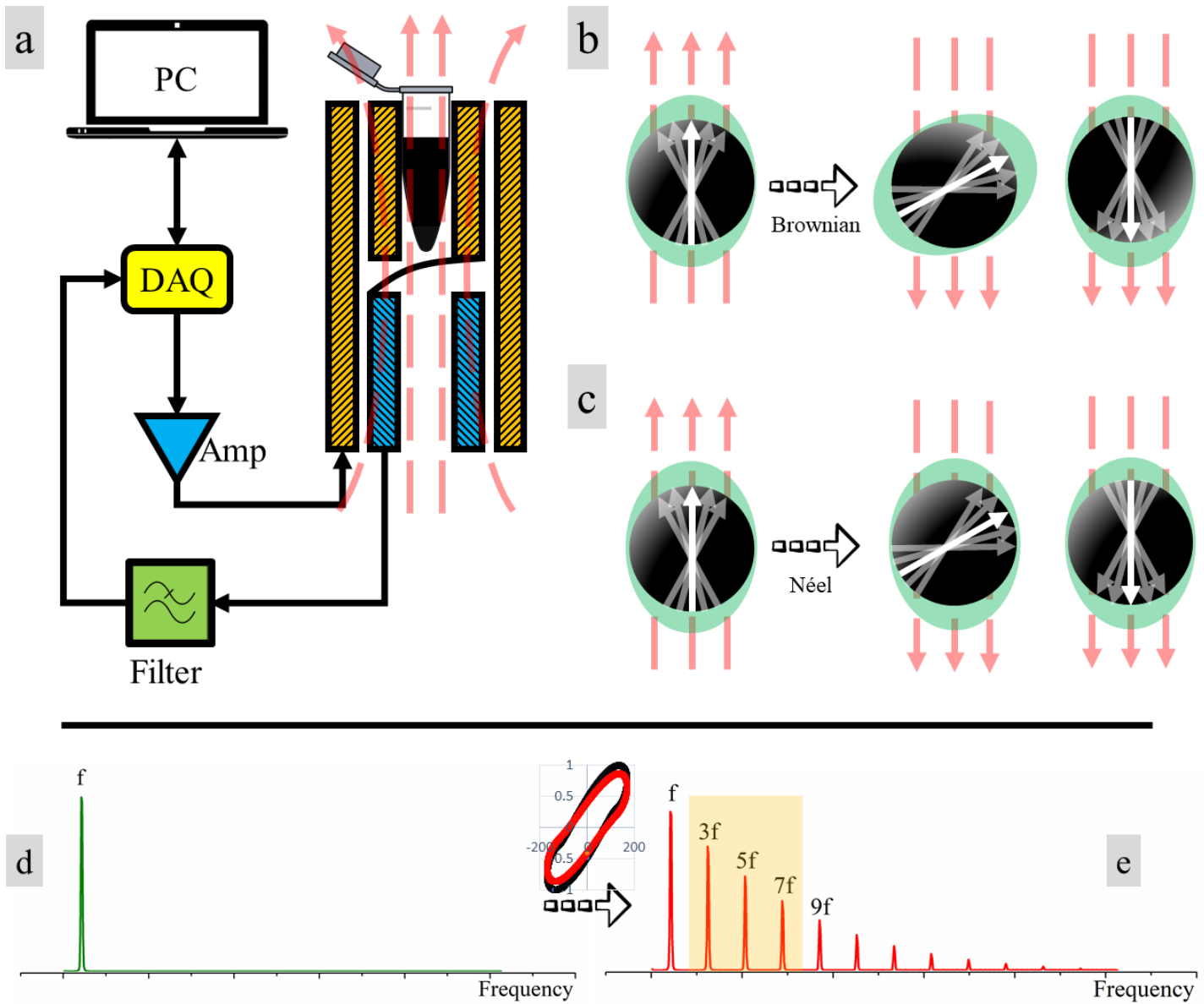

**Figure S1.** (a) Schematic view of homebuilt MPS system setup. (b) Brownian relaxation mechanism of free-rotating MNPs suspended in solution. (c) Néel relaxation mechanism of stationary MNP. (e) and (f) are the spectra from AC driving magnetic field and dynamic magnetic responses containing higher odd harmonics. The inset is the AC/dynamic magnetization curves of γ-$Fe_2O_3$@BM (red) and γ´-$Fe_4N$@BM (black) in OC measured by MPS system.

## S2. Néel relaxation-dominated γ-Fe2O3@BM and γ´-Fe4N@BM MNPs in oleic acid.

For free-rotating MNPs suspended in solution, their magnetic moments relax to align with the externally applied AC magnetic fields through Néel and Brownian relaxations.[12–14] In this section, we investigate the relaxation mechanism of γ-Fe2O3@BM in water and oleic acid (OA).

The Brownian relaxation time is expressed as:

$$\tau_B = \frac{3\eta V_h}{k_B T} \quad (1),$$

where $\eta$ is the viscosity of solution, $k_B$ is the Boltzmann constant, $T$ is temperature, and $V_h$ is the hydrodynamic volume of MNP. Herein, γ-Fe2O3 MNPs with magnetic core diameters $D$ (varied from 1 to 50 nm) and surface OA layer thickness of $c = 2\ nm$ are assumed. Thus, the hydrodynamic volume is expressed as:

$$V_h = \pi(D + 2c)^3/6 \quad (2)$$

The Néel relaxation time is expressed as:

$$\tau_N = \tau_0 e^\sigma \quad (3),$$

$$\sigma = \frac{K_u V_c}{k_B T} \quad (4),$$

Where the typical values for $\tau_0$ are between 10–9 and 10–10 seconds, $K_u$ is the uniaxial anisotropy of γ-Fe2O3@BM, which we used $K_u = 4.6 \times 10^3\ J/m^3$ in this simulation. Magnetic core volume is $V_c = \pi D^3/6$.

It is worth to mention that the Néel and Brownian relaxation models used here are simplified, neglecting the effects of dipolar interactions and magnetic field strength.[5,6,15] Both relaxation processes are dependent on the frequency and amplitude of applied magnetic fields.

The dynamic magnetic responses of MNPs are usually characterized by the effective relaxation time $\tau_{eff}$, which is dependent on Brownian relaxation time $\tau_B$ and Néel relaxation time $\tau_N$. The $\tau_{eff}$ of a free-rotating MNP governs its ability to align its magnetic moment with the external driving field (see Figure S3(a)), this effective relaxation time $\tau_{eff}$ is related to the Brownian and Néel relaxation times and is expressed as:

$$\frac{1}{\tau_{eff}} = \frac{1}{\tau_B} + \frac{1}{\tau_N} \quad (5)$$

The Néel, Brownian, and effective relaxation time of free-rotating γ-Fe2O3@BM is simulated with magnetic core size varied from 1 nm to 50 nm. Figure S2(a) and Figure S2(b) present the relaxation mechanisms of γ-Fe2O3@BM in water (viscosity $\eta = 0.89\ mPa \cdot s$) and oleic acid (viscosity $\eta = 27.64\ mPa \cdot s$) at 25 °C. For γ-Fe2O3@BM dispersed in water (see Figure S2(a)), the critical core size is $D_{crit} = 24\ nm$, where MNPs with core size below 24 nm relax to the external field through a Néel relaxation-dominated process and MNPs with core size above 24 nm relax to the external field through a Brownian relaxation-dominated process. For γ-Fe2O3@BM dispersed in oleic acid (see Figure S2(b)), the critical core size is $D_{crit} = 28\ nm$.

In this paper, we measured the dynamic magnetic responses of γ-Fe$_2$O$_3$@BM in oleic acid, and the average magnetic core size is ~20 nm as shown in the TEM images from Figure 3(a) in the paper. Thus, these γ-Fe$_2$O$_3$@BM MNPs relax to the fast-changing AC magnetic driving field through the Néel relaxation-dominated process.

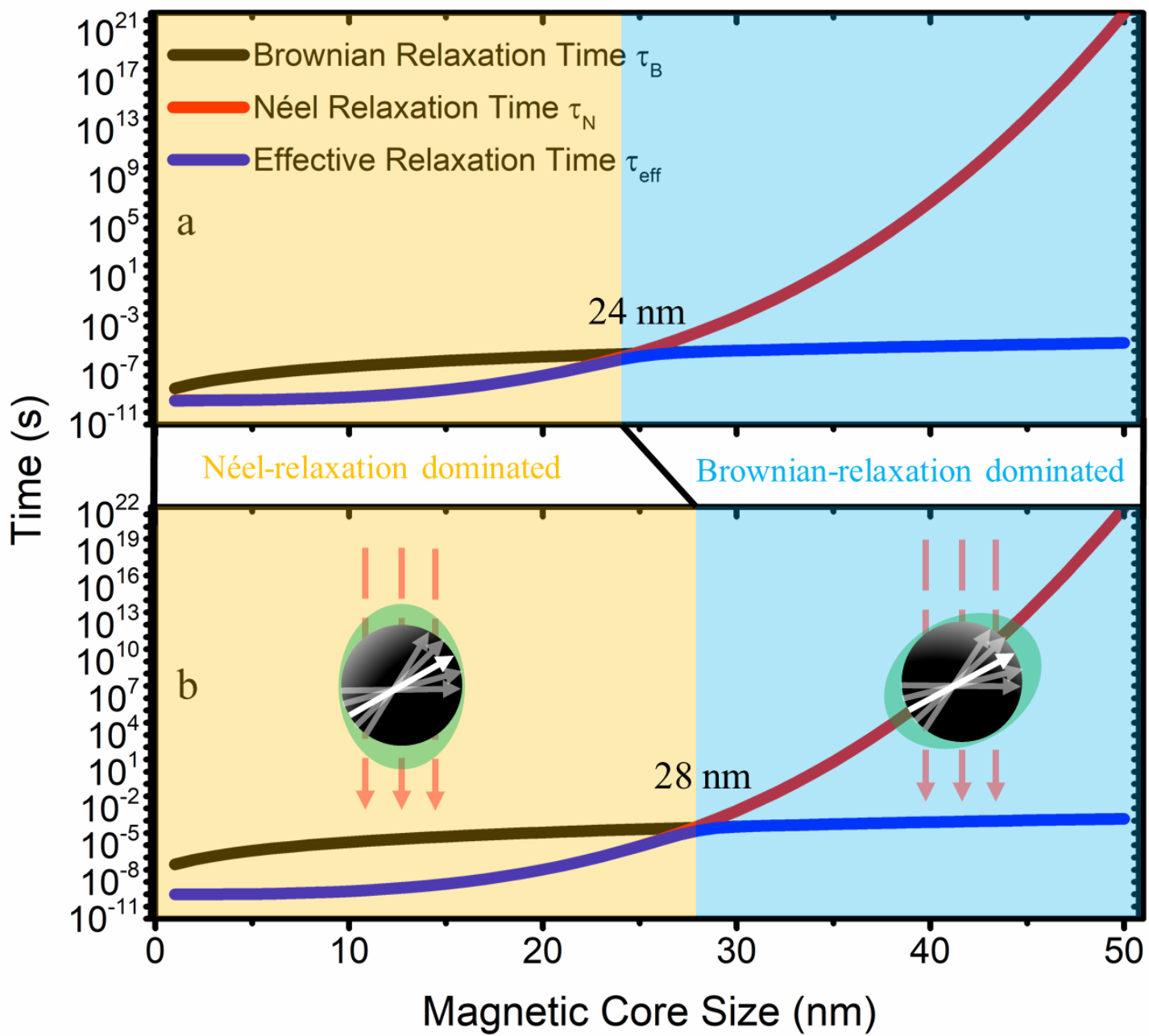

**Figure S2.** Simulated relaxation time of γ-Fe$_2$O$_3$@BM in (a) water and (b) oleic acid at at 25 °C, an oleic acid layer with thickness of 2 nm is assumed on the γ-Fe$_2$O$_3$@BM MNPs.

The relaxation mechanisms of γ′-Fe$_4$N@BM MNPs are more complicated than γ-Fe$_2$O$_3$@BM due to the cubic crystalline anisotropy of γ′-Fe$_4$N@BM MNPs. However, the TEM images from Figure 3(a) in the paper indicate that the Fe MNPs after redox reactions sinter into larger sintered body during the nitridation process due to the high temperature required. Thus, the sintered body, γ´-Fe4N MNPs are around 100 nm in size as confirmed in Figure 3(b), Figure 4(a) and Figure 5 in the paper. As a result, we can safely treat these sintered γ´-Fe4N MNPs as 20 nm γ´-Fe4N MNPs bundling together, in this case, the dominating relaxation mechanism is also Néel relaxation due to that the Brownian relaxation is blocked for these sintered bodies (~100 nm), as shown in Figure S3(b).

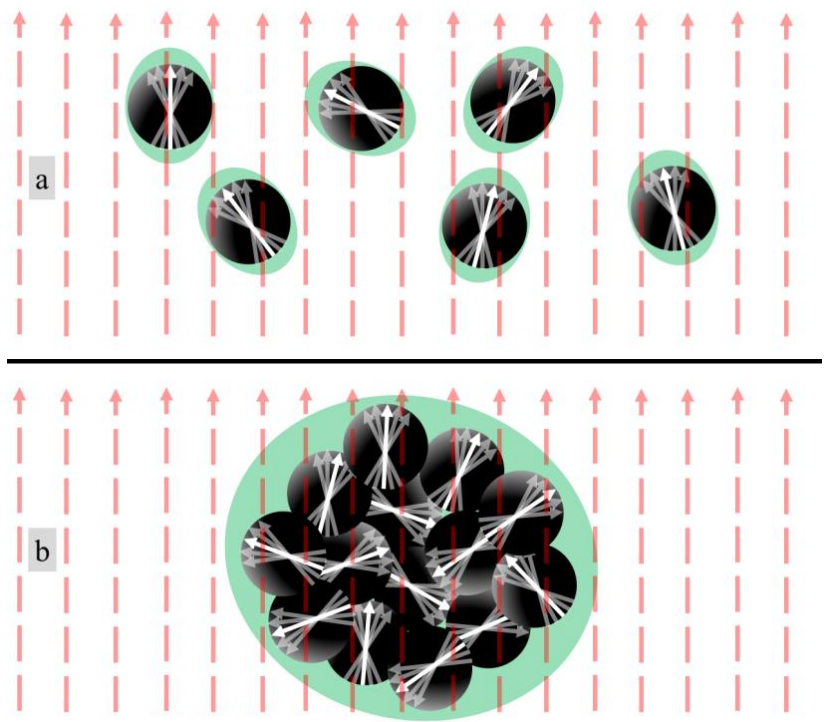

**Figure S3.** (a) Free-rotating γ-Fe2O3@BM MNPs in solution under external magnetic field (red dashed arrow lines). Both Brownian and Néel relaxations are possible for free-rotating MNPs. By choosing γ-Fe2O3@BM MNPs with average magnetic core size of 20 nm, we artificially controlled these γ-Fe2O3@BM MNPs to be Néel relaxation-dominated under AC driving fields. (b) 20 nm γ´-Fe4N MNPs trapped in a sintered body, in solution, under external magnetic field. Brownian relaxation is blocked due to the sintering and only Néel relaxation is possible.

In conclusion, both γ-Fe2O3@BM and γ´-Fe4N@BM MNPs in this work relax to align their magnetic moments to the externally applied AC field through a Néel relaxation-dominated process. The mathematical models used here are assuming perfect spherical MNPs, however, taking the shape anisotropy (irregularly shaped nanoparticles as shown in TEM images from Figure 3 in the paper) and dipolar interactions between MNPs, the practical effective relaxation time is larger than the theoretical calculations.

## S3. Models of dynamic magnetic responses.

In the presence of AC magnetic fields $H(t) = A \cdot sin[2\pi \cdot f \cdot t]$, MNPs are magnetized and their magnetic moments tend to align with the fields. For a monodispersed, non-interacting MNP system, the static magnetic response obeys the Langevin model:

$$M_D(t) = m_s c L\left(\frac{m_s H(t)}{k_B T}\right) \quad (6)$$

where,

$$L(\xi) = \coth \xi - \frac{1}{\xi} \quad (7)$$

The magnetic moment of each MNP is expressed as $m_s = M_s V_c = M_s \pi D^3/6$, where $V_c$ is volume of the magnetic core.

Taylor expansion on $M_D(t)$ shows the major odd harmonic components:

$$\frac{M_D(t)}{m_s c} = L\left(\frac{m_s H(t)}{k_B T}\right)$$

$$= \frac{1}{3}\left(\frac{m_s}{k_B T}\right) H(t) - \frac{1}{45}\left(\frac{m_s}{k_B T}\right)^3 H(t)^3 + \frac{2}{945}\left(\frac{m_s}{k_B T}\right)^5 H(t)^5 + \cdots$$

$$= \cdots + \left[\frac{1}{180} A^3 \left(\frac{m_s}{k_B T}\right)^3 - \frac{1}{1512} A^5 \left(\frac{m_s}{k_B T}\right)^5 + \cdots\right] \times sin[2\pi \cdot 3f \cdot t]$$

$$+ \left[\frac{1}{7560} A^5 \left(\frac{m_s}{k_B T}\right)^5 + \cdots\right] \times sin[2\pi \cdot 5f \cdot t]$$

$$+ \cdots \quad (8)$$

The higher odd harmonics are expressed as:

$$M_D(t)|_{3rd} \approx \frac{m_s c}{180} A^3 \left(\frac{m_s}{k_B T}\right)^3 \times sin[2\pi \cdot 3f \cdot t] \quad (9)$$

$$M_D(t)|_{5th} \approx \frac{m_s c}{7560} A^5 \left(\frac{m_s}{k_B T}\right)^5 \times sin[2\pi \cdot 5f \cdot t] \quad (10)$$

According to the Faraday's law of induction, the induced voltage detected by the pick-up coils is expressed as:

$$u(t) = -S_0 V \frac{d}{dt} M_D(t) \quad (11)$$

where $V$ is volume of MNP suspension. Pick-up coil sensitivity $S_0$ equals to the external magnetic field strength divided by current.

Combining equations (9), (10) and (11), the voltage signal is re-written as:

$$u(t) \propto f \cdot A \cdot m_s \cdot c \cdot V \quad (12)$$

The amplitude of $u(t)$ is proportional to driving field frequency $f$ and amplitude $A$, the magnetic moment $m_s$ of MNPs, the concentration $c$ and volume $V$ of MNP suspension. By effectively controlling the driving field amplitude $A$, the concentration $c$ and volume $V$ of MNP suspension identical. The voltage signal is a function of driving field frequency $f$, the magnetic moment $m_s$ of MNPs. It should be noted that the static magnetic response mode (the Langevin model) discussed above is unable to describe the dynamic magnetic responses of MNPs suspended in solution.[16] Herein, Néel and Brownian relaxation models are introduced in S2 to complete the model.

As shown in Figure 7 from the paper, by controlling the volume and concentration identical (200 μL, 67 mg/mL), the γ´-Fe$_4$N@BM MNP sample show higher harmonic amplitudes over γ-Fe$_2$O$_3$@BM. Which is due to that the γ´-Fe$_4$N@BM MNP has higher saturation magnetization $M_s$ over γ-Fe$_2$O$_3$@BM MNP.

In addition, as discussed in the paper, the harmonic amplitudes of both MNPs increase as the driving field frequency $f$ increases (denoted as region I in Figure 7 from the paper). Within region I, the driving field frequency $f$ is the dominating factor for voltage signal $u(t)$. However, as the AC magnetic field sweeps faster ($f$ increases further), the magnetic moments of MNPs are unable to follow the direction of field, causing larger phase lag $\phi$ between the magnetic moments and field (region III in Figure 7 from the paper). This phase lag attenuates the harmonic amplitudes. The competition between enhancement effect of $f$ and attenuation effect of $\phi$ reaches to a critical point ($f_{crit}$) where the dynamic magnetic responses (harmonic amplitudes) reach to maxima (marked by stars in Figure 7 and Figure 10 from the paper).

## S4. Micromagnetic simulation parameters.

Table S1. Micromagnetic simulation parameters for γ-Fe$_2$O$_3$ MNPs

| Parameters | Description | Values |
|---|---|---|
| **MNP Dimension** | Spherical diameter | (a) 15 nm |
| | | (b) 25 nm |
| | Cubic dimensions | (c) 15 nm × 15 nm × 15 nm |
| | Ellipsoid dimensions | (d) 30 nm × 10 nm × 10 nm |
| **Cell Size** | Length × Width × Thickness | 1 nm × 1 nm × 1 nm |
| $\alpha$ | Gilbert damping factor[17] | 0.2 |
| A | Exchange constant[18] | $10^{-11}$ J/m |
| $M_s$ | Saturation magnetization | 280 kA/m |
| K$_u$ | Uniaxial Anisotropy[19] | 4.6 kJ /m$^3$ |

Table S2. Micromagnetic simulation parameters for γ´-Fe$_4$N MNPs

| Parameters | Description | Values |
|---|---|---|
| **MNP Dimension** | Sintered body, spherical diameter | (e) 100 nm |
| | Ellipsoid dimensions | (f) 200 × 50 nm × 50 nm |
| | Nanoparticle cluster composed of 25 spherical MNPs | (g) 100 nm, each arranged in a 5 by 5 square |
| **Cell Size** | Length × Width × Thickness | 2 nm × 2 nm × 2 nm |
| $\alpha$ | Gilbert damping factor[20] | 1 |
| A | Exchange constant[20] | $15 \times 10^{-12}$ J/m |
| $M_s$ | Saturation magnetization[20] | 1430 × kA/m |
| K$_c$1 | Cubic Anisotropy[20] | $3 \times 10^4$ J/m$^3$ |

## S5. Micromagnetic simulation models

Figure S4(a) – (d) are the evenly dispersed γ-Fe₂O₃ nanoparticles of different shapes and sizes. Figure S4(e) & (f) are the γ´-Fe₄N sintered body models. Figure S4(g) is the γ´-Fe₄N nanoparticle cluster model.

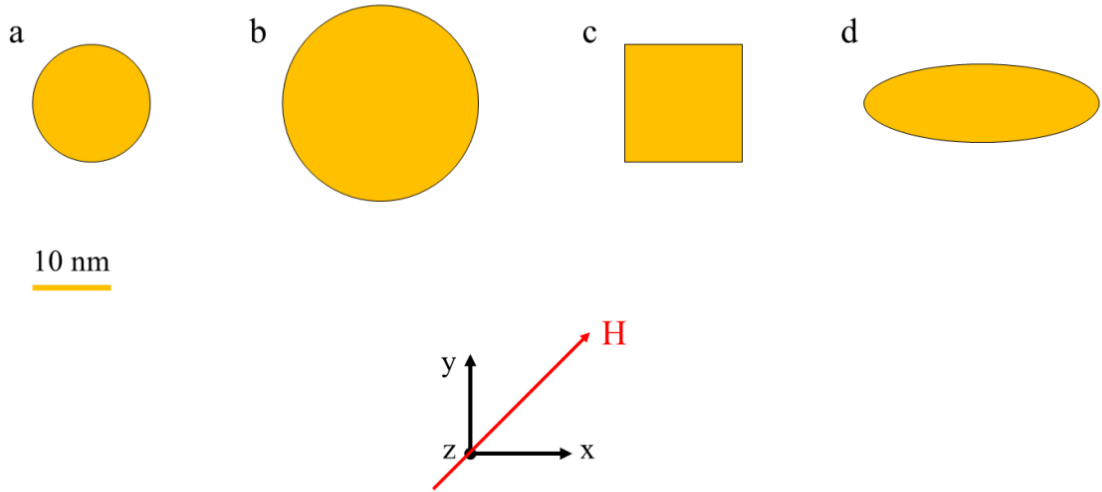

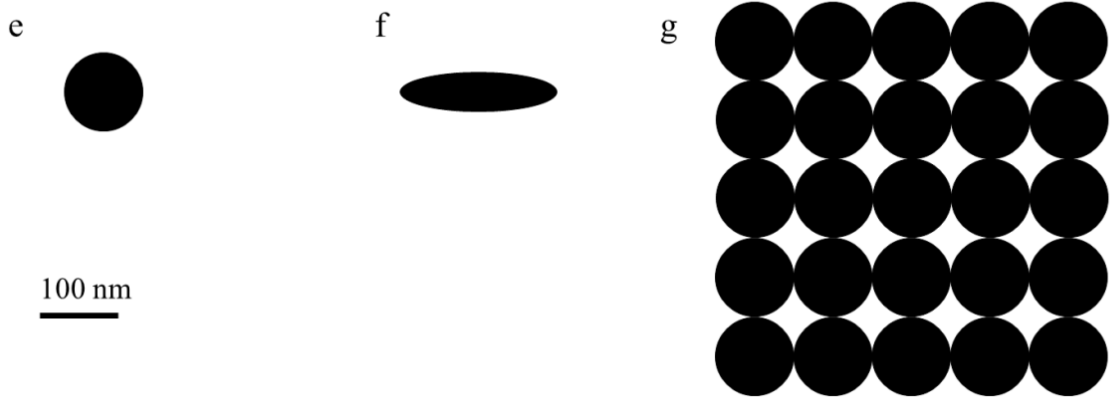

Figure S4. Mumuax3 simulation models in this paper. (a) Spherical γ-Fe₂O₃ nanoparticle with diameter of 15 nm. (b) Spherical γ-Fe₂O₃ nanoparticle with diameter of 25 nm. (c) Cubic γ-Fe₂O₃ nanoparticle with side length of 15 nm. (d) Ellipsoid γ-Fe₂O₃ nanoparticle with long axis of 30 nm and short axis of 10 nm. (e) Sintered body, spherical γ´-Fe₄N nanoparticle with diameter of 100 nm. (f) Sintered body, ellipsoid γ´-Fe₄N nanoparticle with long axis of 200 nm and short axis of 50 nm. (g) γ´-Fe₄N nanoparticle cluster, an array of 5 × 5 spherical γ´-Fe₄N nanoparticle (diameter of 100 nm).

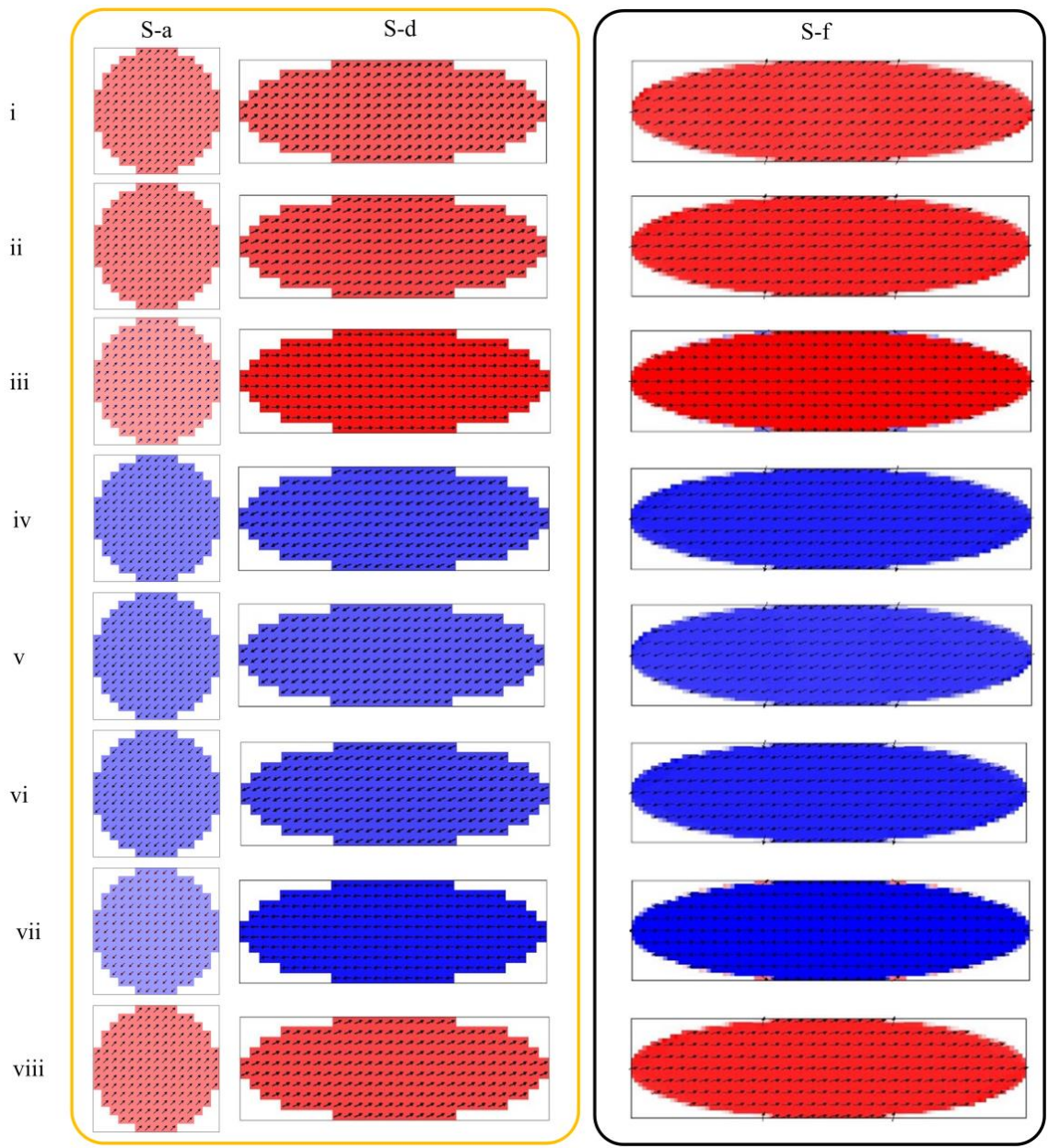

Figure S5. Evolution of magnetizations in different (c) γ-Fe$_2$O$_3$ and (d) γ´-Fe$_4$N nanoparticles under different DC magnetic fields. S-a, S-d, and S-f correspond to the Mumax$_3$ simulation models from Figure S4. S-a: spherical γ-Fe$_2$O$_3$ nanoparticle with diameter of 15 nm; S-d: ellipsoid γ-Fe$_2$O$_3$ nanoparticle with long axis of 30 nm and short axis of 10 nm; S-f: sintered body, ellipsoid γ´-Fe$_4$N nanoparticle with long axis of 200 nm and short axis of 50 nm.

## S6. Magnetocrystalline anisotropy energy of γ-Fe₂O₃ and γ´-Fe₄N.

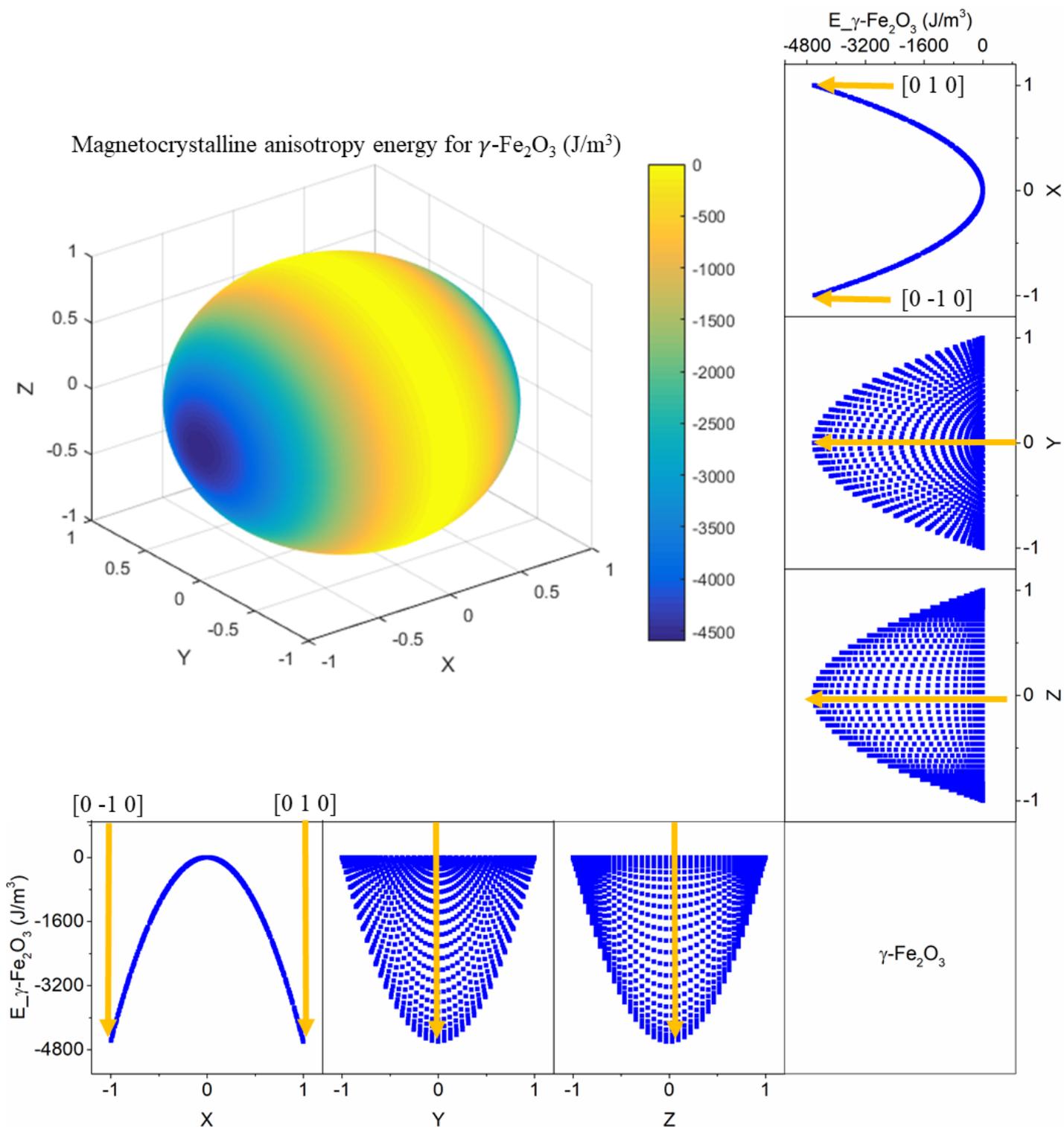

**Figure S6.** Magnetocrystalline anisotropy energy of γ-Fe₂O₃.

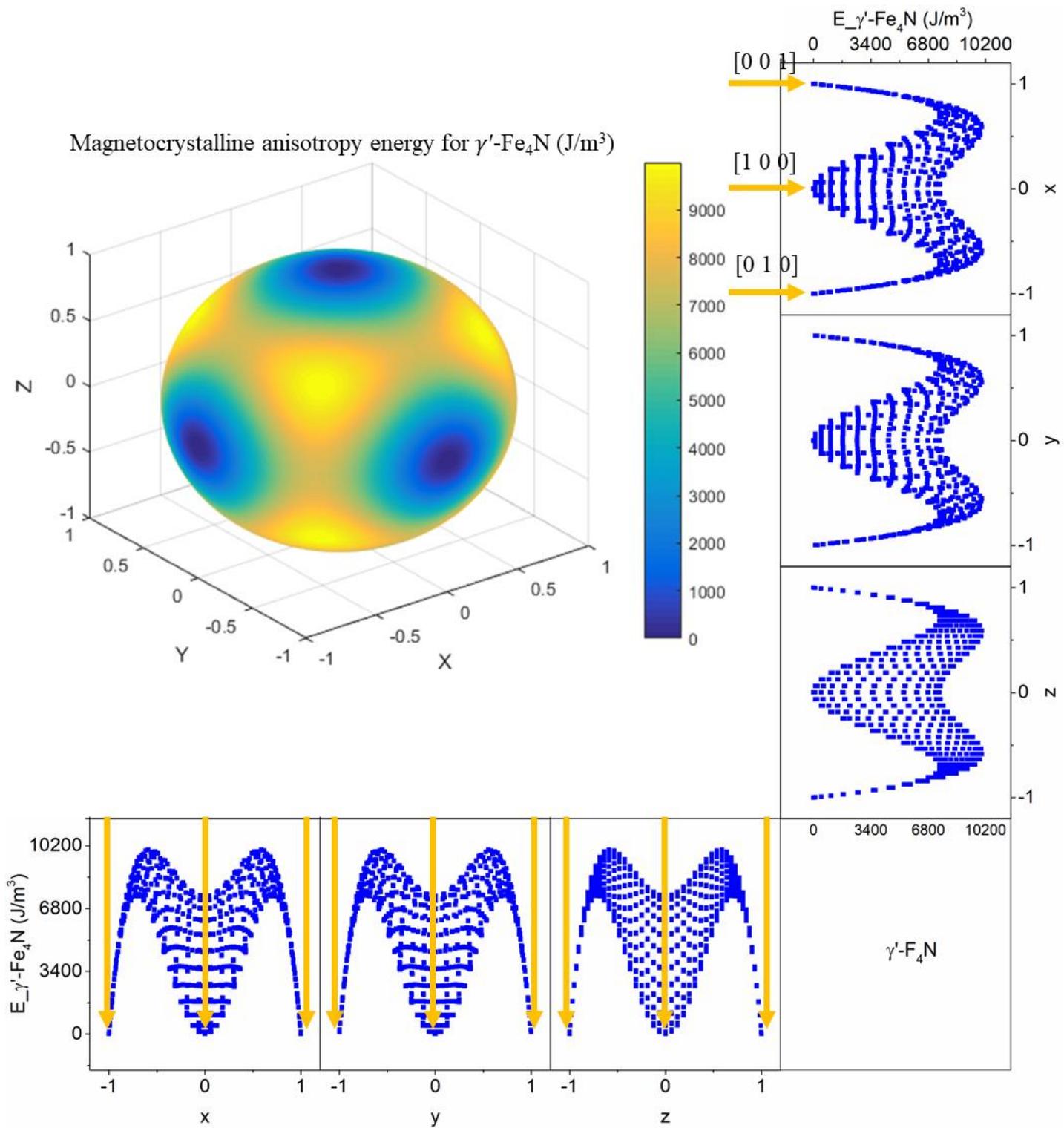

**Figure S7. Magnetocrystalline anisotropy energy of γ´-Fe4N.**

**S7. Stability of γ-Fe₂O₃ and γ´-Fe₄N nanoparticles in oleic acid.**

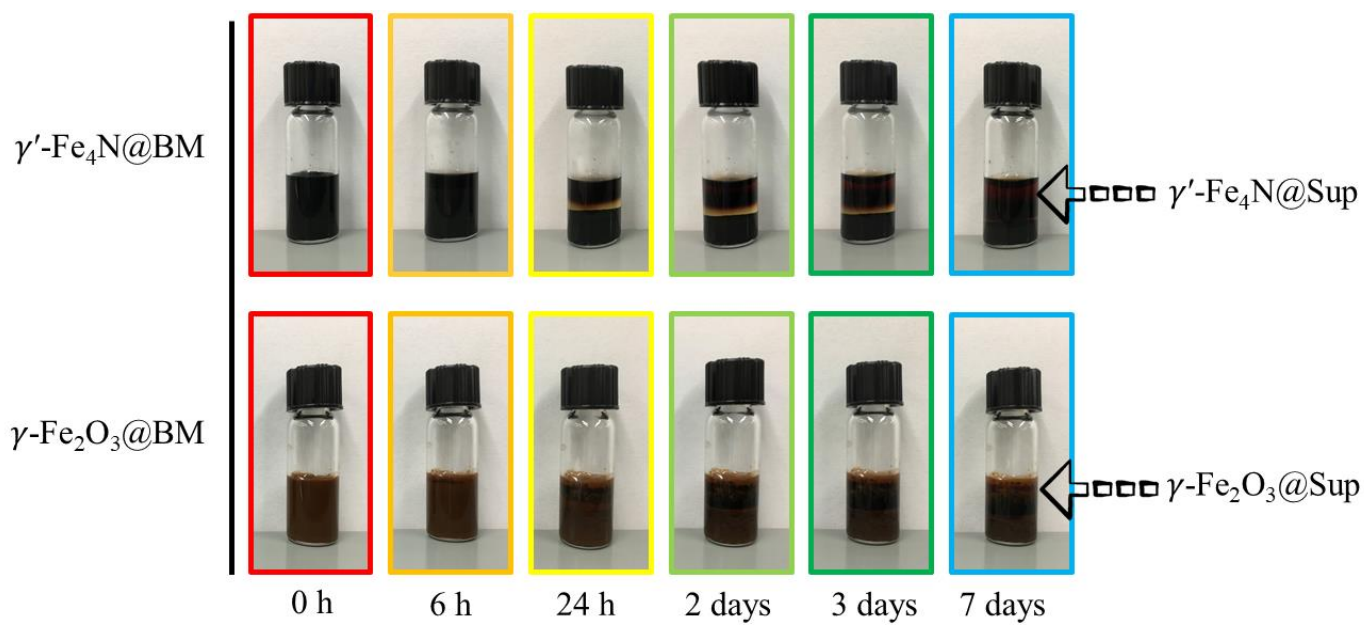

**Figure S8.** Stability of γ-Fe₂O₃ and γ´-Fe₄N nanoparticles in oleic acid.

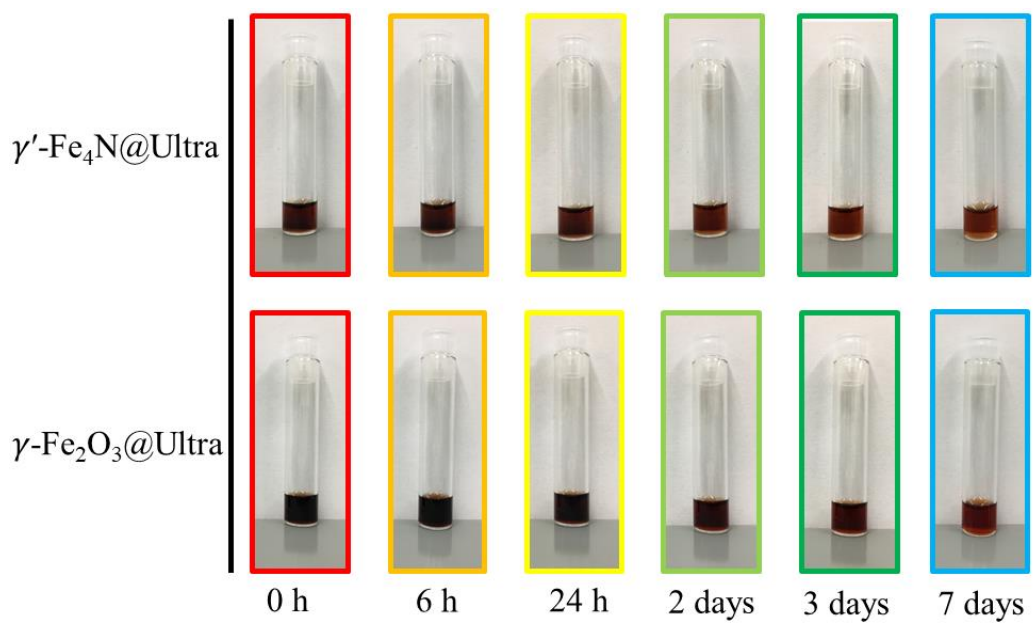

**Figure S9.** Stability of γ-Fe₂O₃@Ultra and γ´-Fe₄N@Ultra samples.

**References**

(1) Achtsnicht, S.; Pourshahidi, A. M.; Offenhäusser, A.; Krause, H.-J. Multiplex Detection of Different Magnetic Beads Using Frequency Scanning in Magnetic Frequency Mixing Technique. *Sensors* **2019**, *19* (11), 2599.

(2) Khurshid, H.; Shi, Y.; Berwin, B. L.; Weaver, J. B. Evaluating Blood Clot Progression Using Magnetic Particle Spectroscopy. *Med. Phys.* **2018**, *45* (7), 3258–3263.

(3) Shah, S. A.; Reeves, D. B.; Ferguson, R. M.; Weaver, J. B.; Krishnan, K. M. Mixed Brownian Alignment and Néel Rotations in Superparamagnetic Iron Oxide Nanoparticle Suspensions Driven by an Ac Field. *Phys. Rev. B* **2015**, *92* (9), 094438.

(4) Zhang, X.; Reeves, D. B.; Perreard, I. M.; Kett, W. C.; Griswold, K. E.; Gimi, B.; Weaver, J. B. Molecular Sensing with Magnetic Nanoparticles Using Magnetic Spectroscopy of Nanoparticle Brownian Motion. *Biosens. Bioelectron.* **2013**, *50*, 441–446.

(5) Wu, K.; Su, D.; Saha, R.; Wong, D.; Wang, J.-P. Magnetic Particle Spectroscopy-Based Bioassays: Methods, Applications, Advances, and Future Opportunities. *J. Phys. Appl. Phys.* **2019**, *52*, 173001.

(6) Wu, K.; Tu, L.; Su, D.; Wang, J.-P. Magnetic Dynamics of Ferrofluids: Mathematical Models and Experimental Investigations. *J. Phys. Appl. Phys.* **2017**, *50* (8), 085005.

(7) Tu, L.; Wu, K.; Klein, T.; Wang, J.-P. Magnetic Nanoparticles Colourization by a Mixing-Frequency Method. *J. Phys. Appl. Phys.* **2014**, *47*, 155001.

(8) Tu, L.; Klein, T.; Wang, W.; Feng, Y.; Wang, Y.; Wang, J.-P. Measurement of Brownian and Néel Relaxation of Magnetic Nanoparticles by a Mixing-Frequency Method. *Magn. IEEE Trans. On* **2013**, *49* (1), 227–230.

(9) Krause, H.-J.; Wolters, N.; Zhang, Y.; Offenhäusser, A.; Miethe, P.; Meyer, M. H.; Hartmann, M.; Keusgen, M. Magnetic Particle Detection by Frequency Mixing for Immunoassay Applications. *J. Magn. Magn. Mater.* **2007**, *311*, 436–444.

(10) Nikitin, P. I.; Vetoshko, P. M.; Ksenevich, T. I. New Type of Biosensor Based on Magnetic Nanoparticle Detection. *J. Magn. Magn. Mater.* **2007**, *311*, 445–449.

(11) Rauwerdink, A. M.; Weaver, J. B. Harmonic Phase Angle as a Concentration-Independent Measure of Nanoparticle Dynamics. *Med. Phys.* **2010**, *37*, 2587–2592.

(12) Viereck, T.; Draack, S.; Schilling, M.; Ludwig, F. Multi-Spectral Magnetic Particle Spectroscopy for the Investigation of Particle Mixtures. *J. Magn. Magn. Mater.* **2019**, *475*, 647–651.

(13) Draack, S.; Lucht, N.; Remmer, H.; Martens, M.; Fischer, B.; Schilling, M.; Ludwig, F.; Viereck, T. Multiparametric Magnetic Particle Spectroscopy of CoFe2O4 Nanoparticles in Viscous Media. *J. Phys. Chem. C* **2019**, *123* (11), 6787–6801.

(14) Orlov, A. V.; Khodakova, J. A.; Nikitin, M. P.; Shepelyakovskaya, A. O.; Brovko, F. A.; Laman, A. G.; Grishin, E. V.; Nikitin, P. I. Magnetic Immunoassay for Detection of Staphylococcal Toxins in Complex Media. *Anal. Chem.* **2013**, *85*, 1154–1163.

(15) Wu, K.; Su, D.; Saha, R.; Liu, J.; Wang, J.-P. Investigating the Effect of Magnetic Dipole-Dipole Interaction on Magnetic Particle Spectroscopy (MPS): Implications for Magnetic Nanoparticle-Based Bioassays and Magnetic Particle Imaging (MPI). *J. Phys. Appl. Phys.* **2019**.

(16) Reeves, D. B.; Weaver, J. B. Combined Néel and Brown Rotational Langevin Dynamics in Magnetic Particle Imaging, Sensing, and Therapy. *Appl. Phys. Lett.* **2015**, *107* (22), 223106.

(17) Draine, B. T.; Hensley, B. Magnetic Nanoparticles in the Interstellar Medium: Emission Spectrum and Polarization. *Astrophys. J.* **2013**, *765* (2), 159.

(18) Wu, W.; Xiao, X.; Zhang, S.; Peng, T.; Zhou, J.; Ren, F.; Jiang, C. Synthesis and Magnetic Properties of Maghemite (γ-Fe$_2$O$_3$) Short-Nanotubes. *Nanoscale Res. Lett.* **2010**, *5* (9), 1474.

(19) Komorida, Y.; Mito, M.; Deguchi, H.; Takagi, S.; Millán, A.; Silva, N.; Palacio, F. Surface and Core Magnetic Anisotropy in Maghemite Nanoparticles Determined by Pressure Experiments. *Appl. Phys. Lett.* **2009**, *94* (20), 202503.

(20) Ito, K.; Rougemaille, N.; Pizzini, S.; Honda, S.; Ota, N.; Suemasu, T.; Fruchart, O. Magnetic Domain Walls in Nanostrips of Single-Crystalline Fe4N (001) Thin Films with Fourfold in-Plane Magnetic Anisotropy. *J. Appl. Phys.* **2017**, *121* (24), 243904.